\documentclass[fleqn,usenatbib]{mnras}

\usepackage{newtxtext,newtxmath}

\usepackage[T1]{fontenc}

\DeclareRobustCommand{\VAN}[3]{#2}
\let\VANthebibliography\thebibliography
\def\thebibliography{\DeclareRobustCommand{\VAN}[3]{##3}\VANthebibliography}

\usepackage{graphicx}	
\usepackage{amsmath}	
\usepackage{mathtools}

\usepackage{color}
\usepackage[normalem]{ulem}  
\usepackage{xspace}
\usepackage{xcolor}
\definecolor{green2}{rgb}{0,0.5,0}
\definecolor{green2}{rgb}{0.2,0.6,0.1}
\definecolor{orange}{rgb}{1,0.5,0}

\title[HMNS Nucleosynthesis]{$r$-process Nucleosynthesis and Kilonovae from Hypermassive Neutron Star Remnants}

\author[S. Curtis et al.]{Sanjana Curtis$^{1}$\thanks{E-mail: s.sanjana@uva.nl},
Philipp M\"osta$^{2}$,
Zhenyu Wu$^{3}$, 
David Radice$^{4,5,6}$, Luke Roberts$^{7}$,
Giacomo Ricigliano $^{8}$,
\newauthor{and Albino Perego $^{8,9}$}
\\
$^{1}$Anton Pannekoek Institute for Astronomy, University of Amsterdam, Science Park 904, 1098 XH Amsterdam, The Netherlands\\
$^{2}$GRAPPA, Anton Pannekoek Institute for Astronomy and Institute of
High-Energy Physics, University of Amsterdam,\\
Science Park 904, 1098 XH Amsterdam, The Netherlands\\
$^{3}$ School of Astronomy and Space Science, Nanjing University, Nanjing 210023, China\\
$^{4}$ Institute for Gravitation and the Cosmos, The Pennsylvania State University, University Park, PA 16802, USA \\
$^{5}$ Department of Physics, The Pennsylvania State University, University Park, PA 16802, USA \\
$^{6}$ Department of Astronomy \& Astrophysics, The Pennsyvlania State University, University Park, PA 16802, USA \\
$^{7}$ Computer, Computational, and Statistical Sciences Division, Los Alamos National Laboratory, Los Alamos, NM, 87545, USA\\
$^{8}$ Dipartimento di Fisica, Università di Trento, Via Sommarive 14, 38123 Trento, Italy\\
$^{9}$ INFN-TIFPA, Trento Institute for Fundamental Physics and Applications, ViaSommarive 14, I-38123 Trento, Italy
}
\graphicspath{{./}{figures/}}
\date{Accepted XXX. Received YYY; in original form ZZZ}

\pubyear{2015}

\begin{document}
\label{firstpage}
\pagerange{\pageref{firstpage}--\pageref{lastpage}}
\maketitle

\begin{abstract}
We investigate $r$-process nucleosynthesis and kilonova emission resulting from binary neutron star (BNS) mergers based on a three-dimensional (3D) general-relativistic magnetohydrodynamic (GRMHD) simulation of a hypermassive neutron star (HMNS) remnant. The simulation includes a microphysical finite-temperature equation of state (EOS) and neutrino emission and absorption effects via a leakage scheme. We track the thermodynamic properties of the ejecta using Lagrangian tracer particles and determine its composition using the nuclear reaction network \texttt{SkyNet}. We investigate the impact of neutrinos on the nucleosynthetic yields by varying the neutrino luminosities during post-processing. The ejecta show a broad distribution with respect to their electron fraction $Y_e$, peaking between $\sim$0.25-0.4 depending on the neutrino luminosity employed. We find that the resulting $r$-process abundance patterns differ from solar, with no significant production of material beyond the second $r$-process peak when using luminosities recorded by the tracer particles. We also map the HMNS outflows to the radiation hydrodynamics code \texttt{SNEC} and predict the evolution of the bolometric luminosity as well as broadband light curves of the kilonova. The bolometric light curve peaks on the timescale of a day and the brightest emission is seen in the infrared bands. This is the first direct calculation of the $r$-process yields and kilonova signal expected from HMNS winds based on 3D GRMHD simulations. For longer-lived remnants, these winds may be the dominant ejecta component producing the kilonova emission.

\end{abstract}




\section{Introduction}

The inspiral and merger of binary neutron stars is accompanied by the ejection of neutron-rich matter that undergoes rapid neutron capture ($r$-process) nucleosynthesis \citep{Lattimer1977, Symbalisty1982, Meyer1989, Goriely2011, Cowan2021}. Radioactive decay of unstable $r$-process nuclei synthesized in these ejecta can power an electromagnetic transient called a kilonova \citep{LiandPac1998, Metzger2010, Kasen2013}. In August 2017, gravitational waves from the merger of a pair of neutron stars were detected for the first time along with their kilonova counterpart: GW170817 \citep{Abbott2017} and AT2017gfo \citep{Coulter2017, SoaresSantos2017, Arcavi2017}, respectively. The electromagnetic spectrum of this kilonova provided the first direct evidence that such mergers are a site where $r$-process nucleosynthesis takes place and produces the heaviest elements in our Universe \citep{Kasen2017, Pian2017}. The observed kilonova emission started out with a featureless thermal spectrum that peaked at UV/optical frequencies \citep{Evans2017, McCully2017, Nicholl2017}, rapidly evolving over the next few days to show a spectral peak in the near-infrared \citep{Pian2017, Tanvir2017, Chornock2017}. This behavior of the kilonova associated with GW170817 is usually interpreted as resulting from distinct ejecta components, giving rise to the early blue and late red peaks \citep{Cowperthwaite2017, Drout2017}. Relatively neutron-rich ejecta that synthesize a substantial mass fraction of lanthanides, which are opaque to blue light, produce a ``red'' kilonova, while the ``blue'' kilonova is understood as emission arising from a less neutron-rich, lanthanide-poor component of the ejecta with a correspondingly lower opacity.  

The total ejecta mass inferred for the red kilonova component is  $M_{\rm{red}} \approx$ 5 x 10$^{-2} M_{\odot}$, with an electron fraction $Y_e \lesssim$ 0.2 and moving at a mean velocity of $v_{\rm{red}} \approx 0.1c$ \citep{Villar2017}. The dynamical merger ejecta and/or outflows from the remnant accretion torus, the latter possibly being the dominant component, can provide the red kilonova ejecta. The origin of the blue kilonova is less understood. The ejecta properties derived for the blue component include a total mass $M_{\rm{blue}} \approx$ 2 x 10$^{-2} M_{\odot}$, high velocities $v_{\rm{blue}} \approx$ 0.2 - 0.3$c$ and a relatively high $Y_e \approx$ 0.25-0.35. Shock-heated polar dynamical ejecta can produce high-velocity outflows with the requisite $Y_e$ but cannot account for the large quantity of mass needed to explain the blue kilonova.
Neutrino-driven winds from an HMNS remnant offer another possible explanation since they experience enhanced neutrino reprocessing that drives up the $Y_e$. However, most simulations so far have found mass-averaged velocities no larger than $\sim 0.1c$ for these winds \citep{Fahlman2018}, much slower than inferred for the blue component. An alternative explanation was suggested by \cite{Metzger2018} who proposed that neutrino-heated, magnetically-accelerated winds from a strongly magnetized HMNS remnant could simultaneously provide the high total mass, high velocity, as well as the relatively high $Y_e$ needed to produce a blue kilonova. Another possible mechanism powering the blue kilonova was suggested in \cite{Nedora2019}, who found that spiral density waves in the remnant generate a wind of mass $\sim$10$^{-2} M_{\odot}$, velocity $\sim 0.2c$, and typical ejecta $Y_e$ above 0.25.

Recently, \cite{Moesta2020} have carried out high-resolution 3D dynamical GRMHD simulations where they evolve a post-merger HMNS remnant with an initial poloidal magnetic field. They found a magnetized neutron-rich wind driven from the HMNS, which ejects material at a rate of $\sim$ 0.1 $M_{\odot}$ s$^{-1}$ resulting in a total ejecta mass of 5 $\times$ 10$^{-3} M_{\odot}$. These HMNS ejecta thus represent an important component of BNS ejecta in addition to the dynamical ejecta (10$^{-4}$ $M_{\odot} <$ M$_{\rm{ej}} <$ 10$^{-2}$ $M_{\odot}$) and winds driven from the accretion disk once a black hole (BH) has formed. For longer-lived remnants, they could be the dominant ejecta component. The magnetized outflows also have a broad distribution in velocity space, with a significant fraction of the material with velocities in the range of $0.3c \lesssim v^r \lesssim 0.5c$, setting these ejecta apart from the dynamical ejecta ($v^r < 0.3c$) and the accretion disk winds ($v^r < 0.1c$).

In this work, we post-process the HMNS outflow to track the $r$-process abundances in the ejecta. We calculate abundances using different constant neutrino luminosities during post-processing in addition to using luminosities recorded by the tracer particles. This allows us to constrain the impact of uncertainties introduced by the approximate neutrino leakage scheme employed in the dynamical simulation. We map the outflow to a spherically-symmetric radiation hydrodynamics code and track its further evolution to predict the bolometric light curve of the resulting kilonova. We also produce light curves in different bands under the assumption of blackbody emission. We find that these ejecta do not produce a robust $r$-process up to the third $r$-process peak. The bolometric light curve of the kilonova evolves quickly and reaches a peak luminosity of $\sim$10$^{41}$ erg s$^{-1}$ at around one day, with the brightest emission observed in the infrared bands. 

The paper is organized as follows. In Section \ref{sec:methods}, we describe our input models, mapping procedure and numerical codes used to compute abundances and light curves. In Section \ref{sec:results_abund}, we present the ejecta composition followed by the bolometric and broadband light curves of the kilonova in Section \ref{sec:results_kn}. We discuss the implications of our results and future directions in Section \ref{sec:summary}.  

\section{Methods}
\label{sec:methods}
\subsection{Input Models}

We study model B15-low, presented in \cite{Moesta2020} where an HMNS post-merger remnant is evolved with an initial poloidal magnetic field of strength 10$^{15}$ G. The simulation employs ideal GRMHD using the \texttt{Einstein Toolkit} and includes the $K_0 = 220$ MeV variant of the equation of state of \cite{Lattimer1991}. Neutrinos are treated via an approximate leakage/heating scheme that captures the overall energetics and lepton number exchange due to neutrino emission and absorption \citep{OConnor2010,Ott2013}. This scheme tracks three neutrino species: electron flavor neutrinos $\nu_e$, electron flavor antineutrinos $\bar{\nu}_e$, and all the heavy-lepton flavor neutrinos grouped together into a single species called $\nu_x$. The scheme captures the overall neutrino energetics correctly up to a factor of a few compared to full neutrino transport in simulations of core-collapse supernovae \citep{OConnor2010}. It does not account for momentum deposition, energy dependence, or neutrino pair-annihilation.

The HMNS evolved using the setup described above was formed in the merger of an equal-mass binary with individual NS masses of 1.35$M_{\odot}$ at infinity, originally simulated in GRHD with the \texttt{WhiskyTHC} code in \cite{Radice2018}. It is mapped as initial data at 17 ms post-merger, adding a poloidal magnetic field of strength $B_0$=10$^{15}$ G. 

During the course of the simulation, this initial magnetic field is amplified due to MRI-induced turbulence in the HMNS to magnetar strengths. A magnetized neutron-rich wind is driven from the HMNS, which ejects material at a rate of $\sim$ 4.6 $\times$ 10$^{-2}$ $M_{\odot}$s$^{-1}$ and accounts for the majority of the ejected mass. Magnetic fields collimate part of the outflow into a mildly relativistic jet. The total ejecta mass estimated from the low-resolution simulation is 1.1 $\times 10^{-3} M_{\odot}$, making these ejecta an important component in BNS mergers for both $r$-process nucleosynthesis and the resulting kilonova. The outflow persists until the HMNS collapses to a black hole $\sim$21 ms after the start of the simulation i.e. $\sim$38 ms post-merger. 

\begin{figure*}
\begin{center}
        \begin{tabular}{cccc}
        \includegraphics[width=0.24\textwidth]{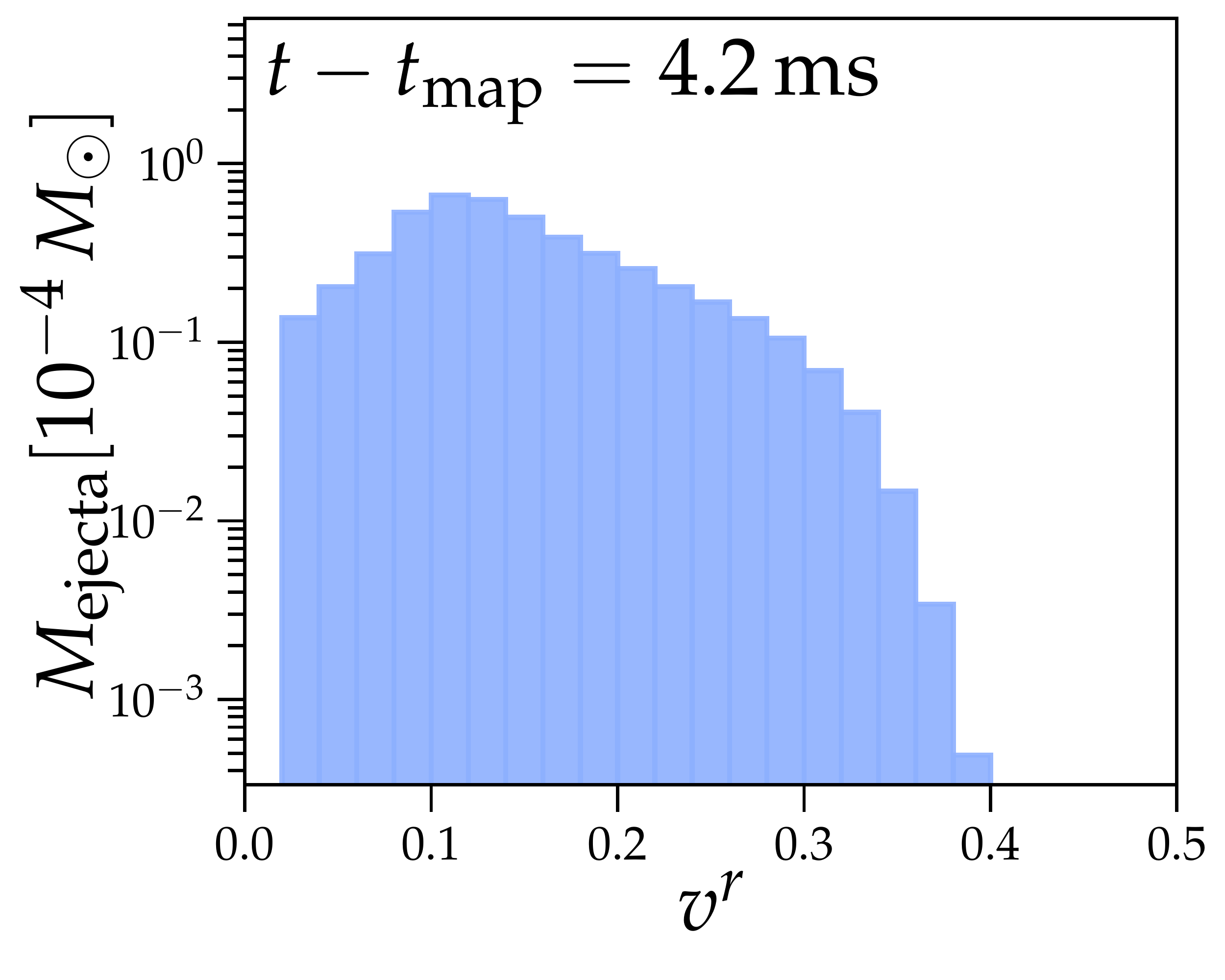}
        \includegraphics[width=0.24\textwidth]{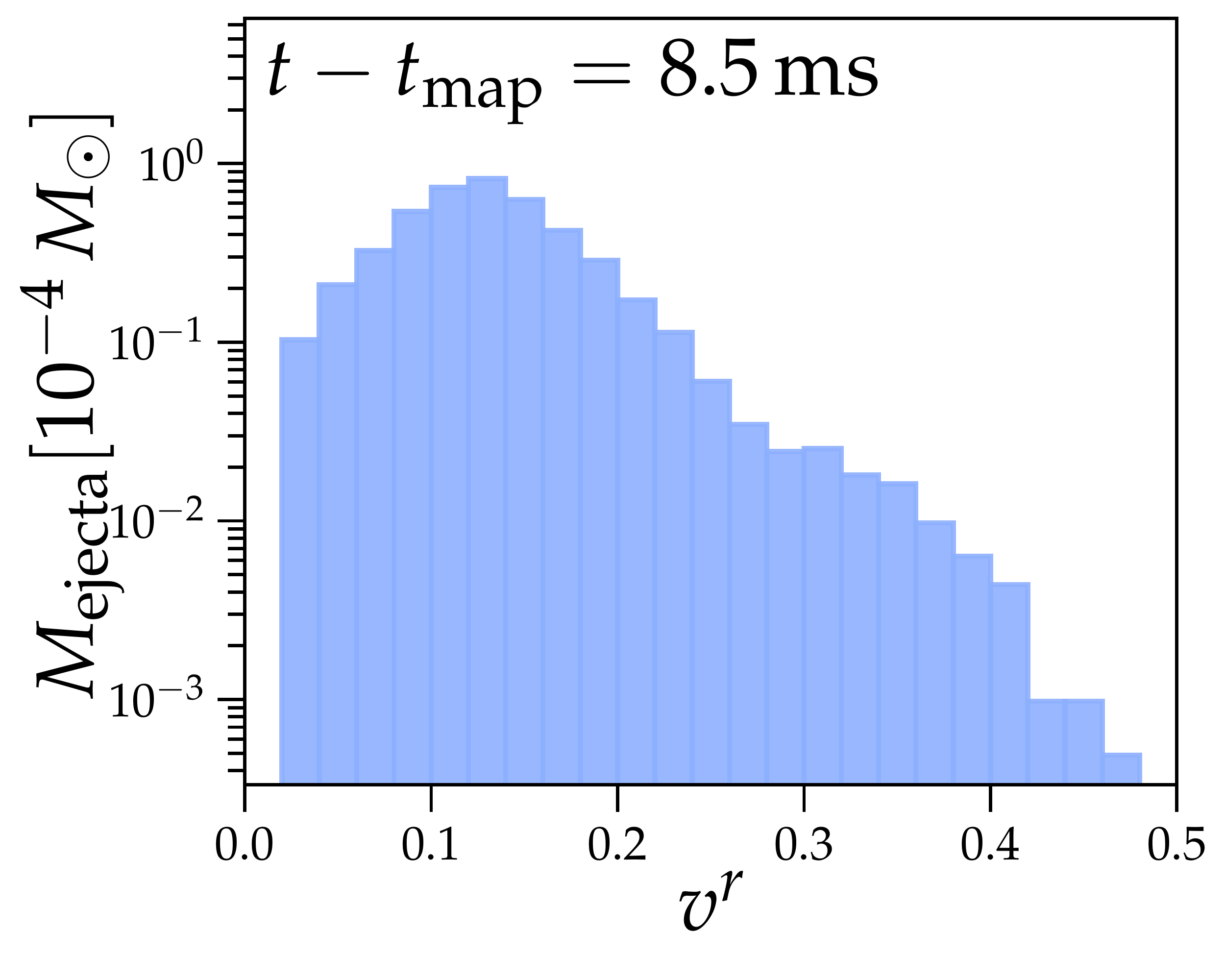}
        \includegraphics[width=0.24\textwidth]{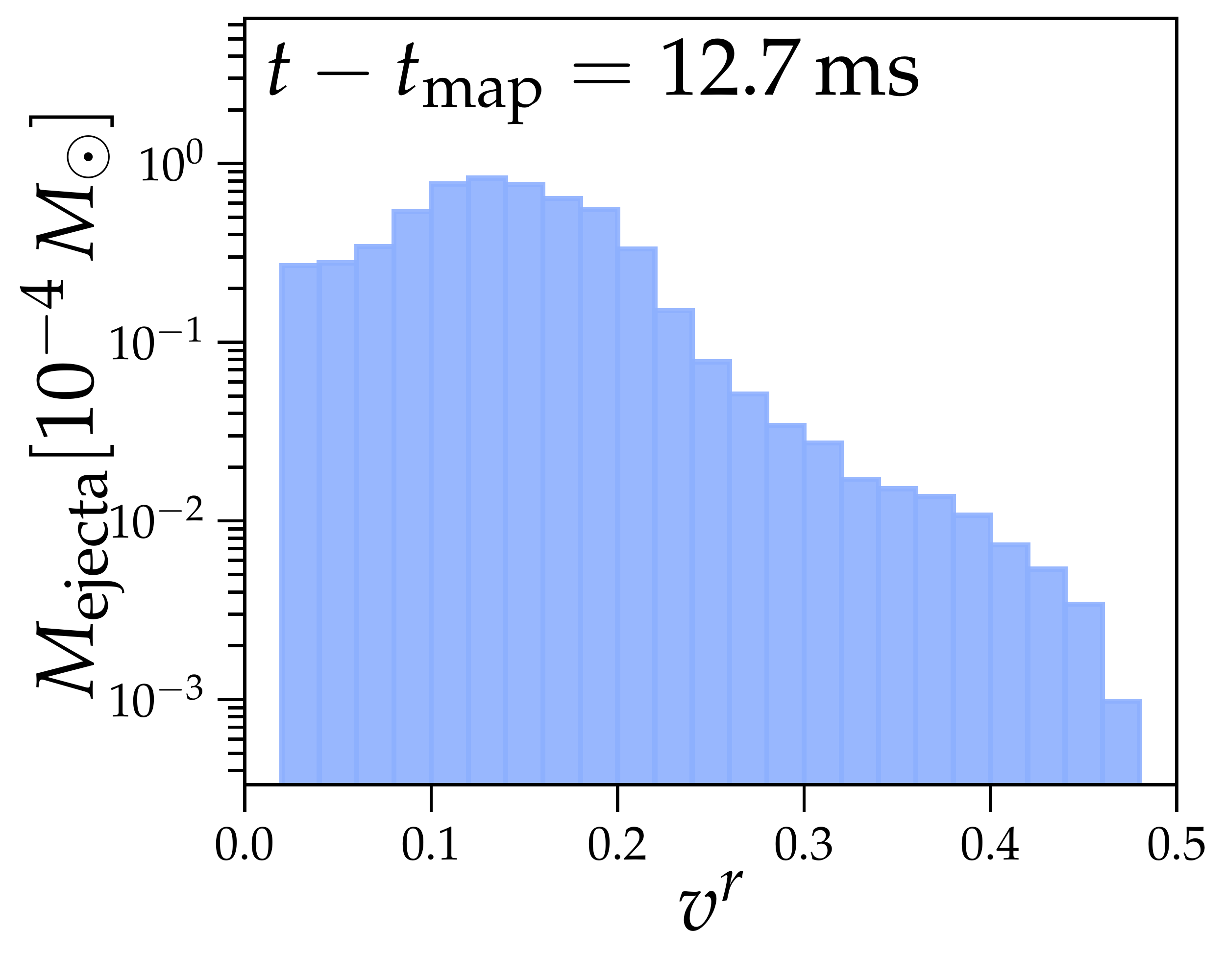}
        \includegraphics[width=0.24\textwidth]{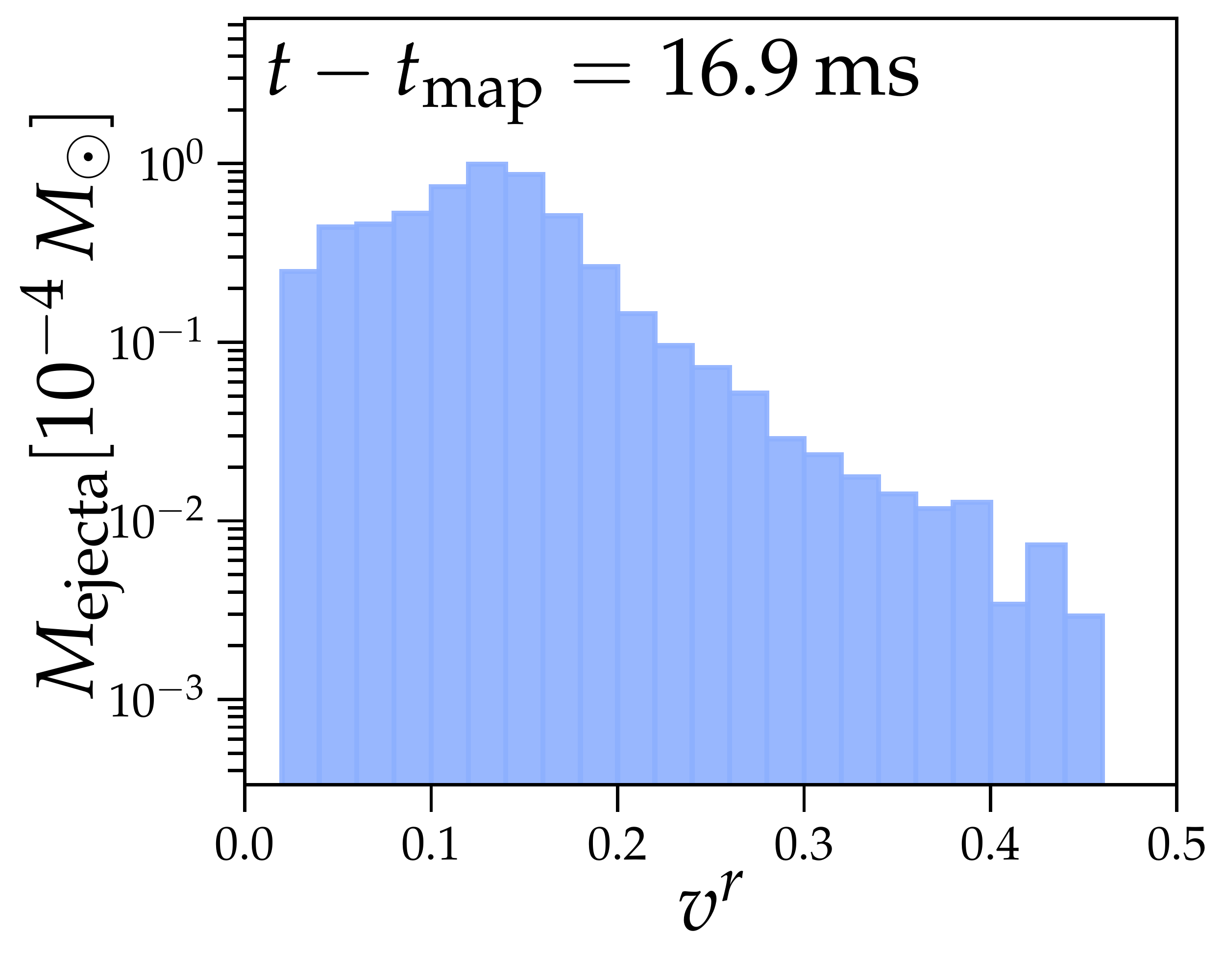}\\
        \includegraphics[width=0.24\textwidth]{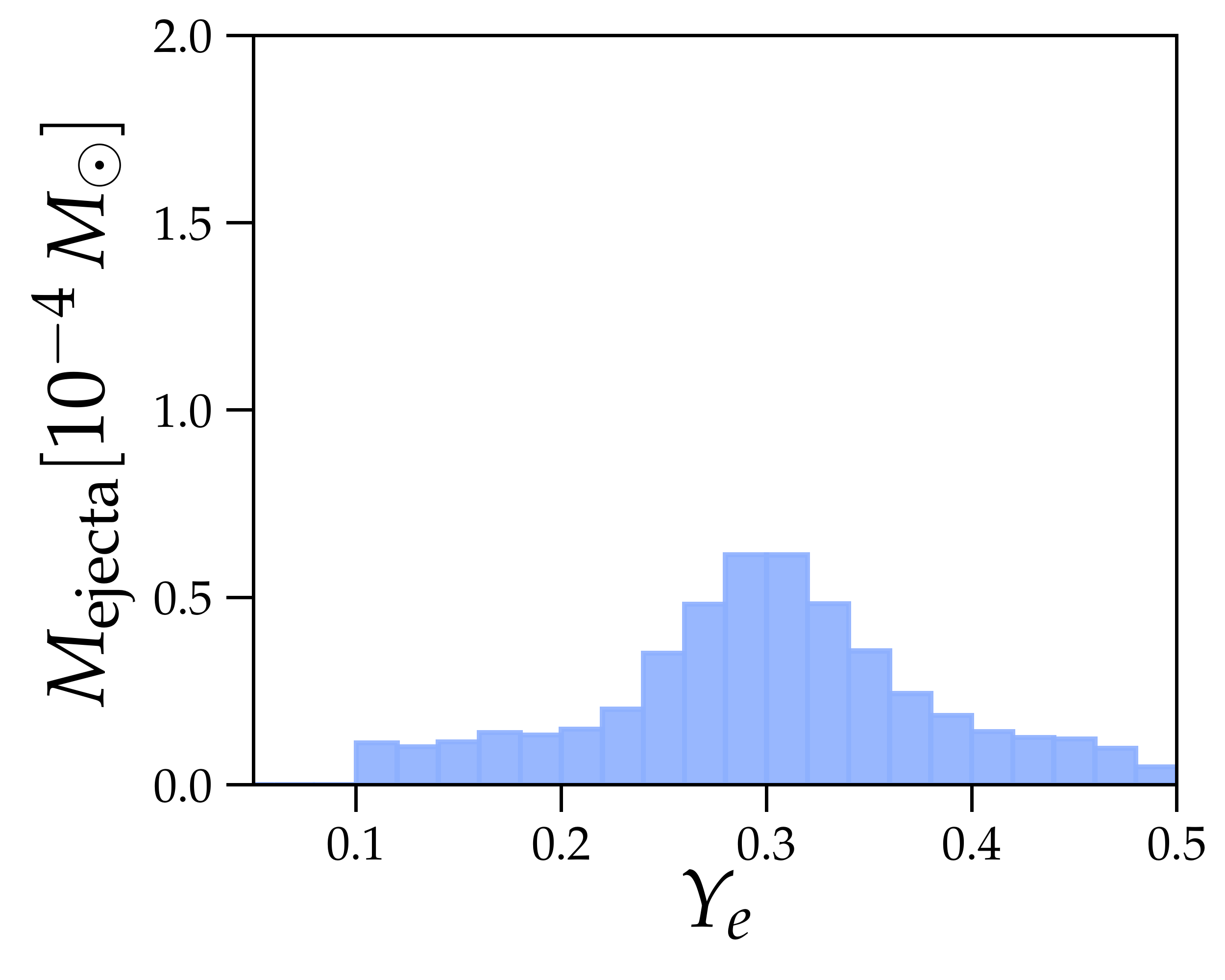}
        \includegraphics[width=0.24\textwidth]{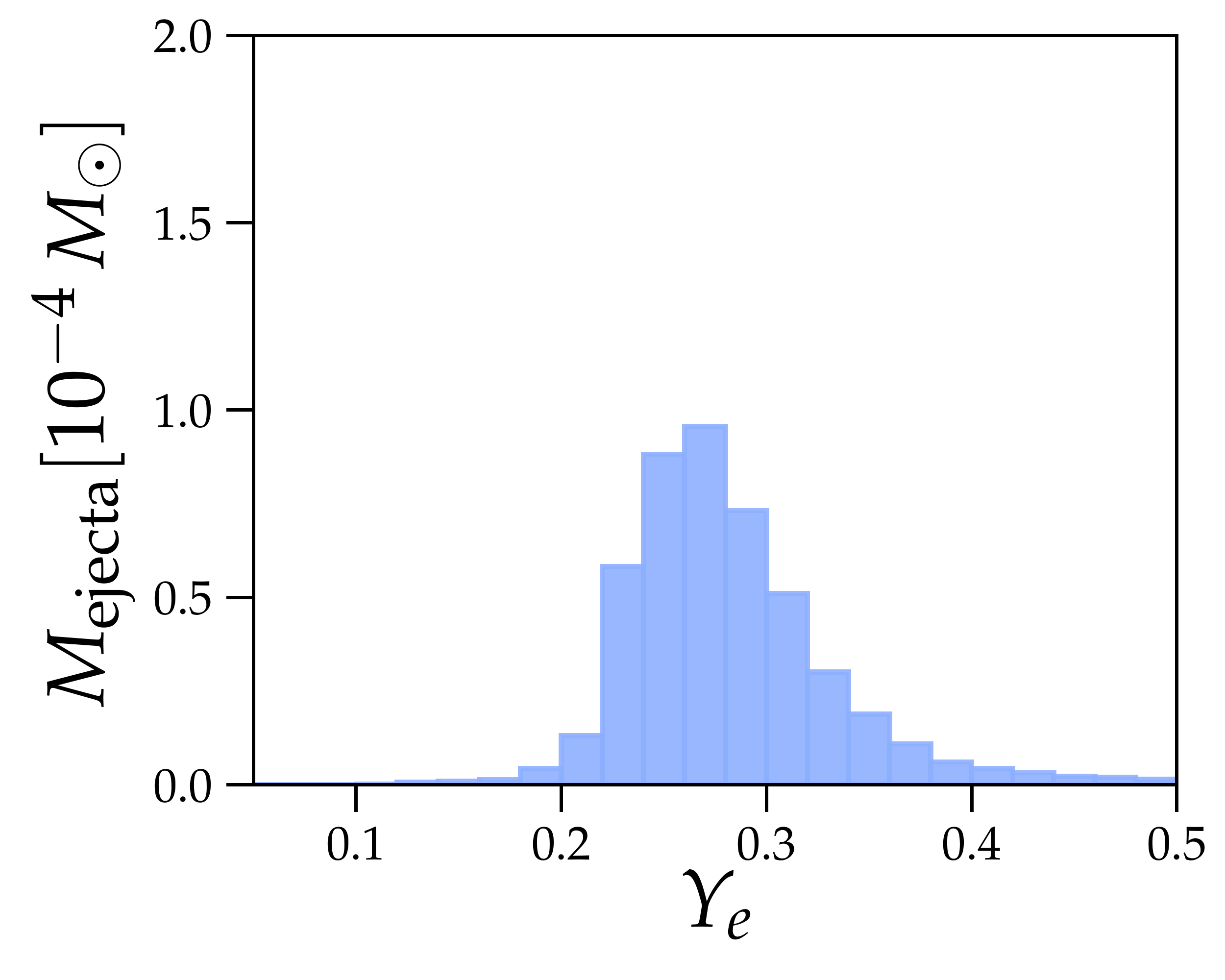}
        \includegraphics[width=0.24\textwidth]{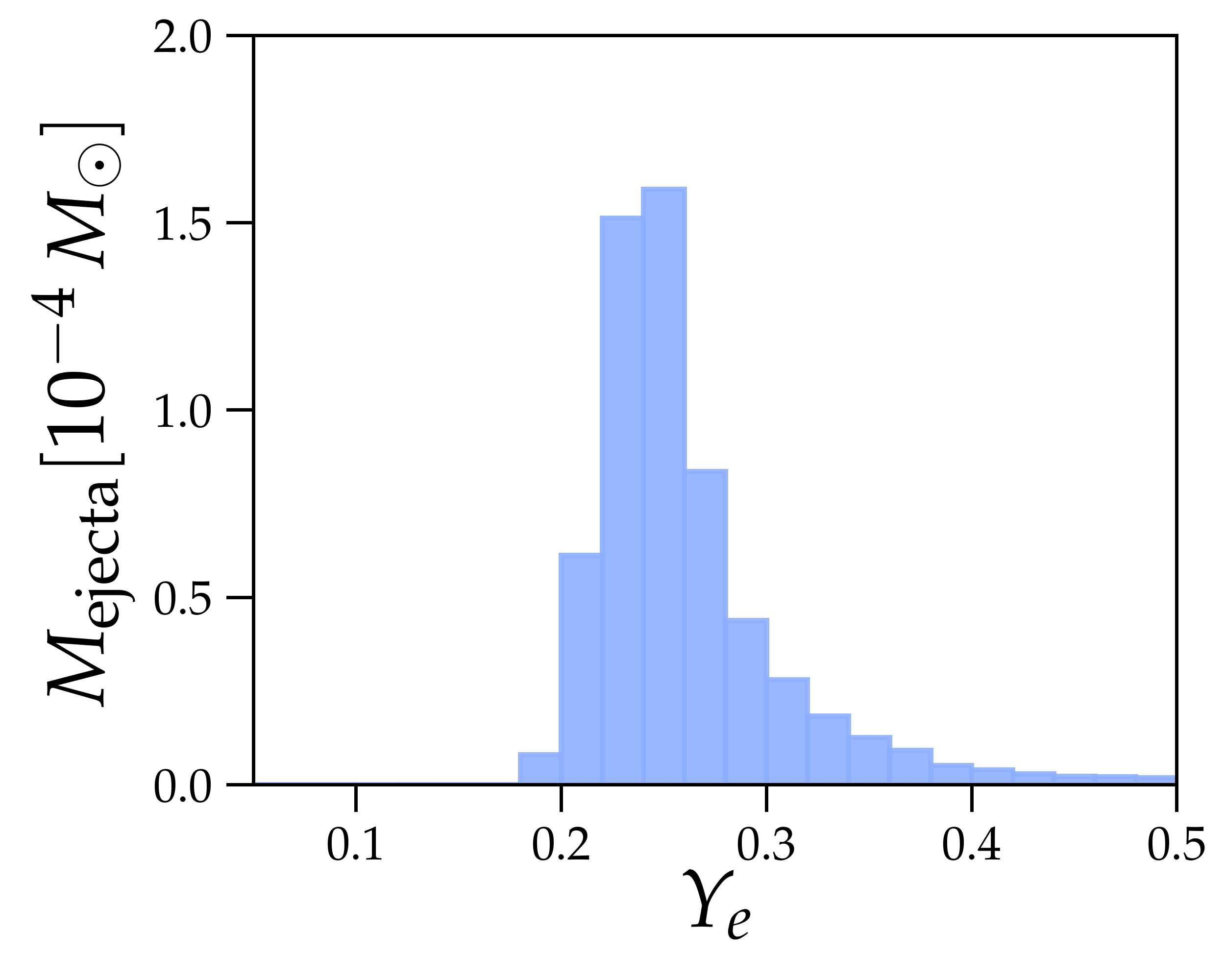}
        \includegraphics[width=0.24\textwidth]{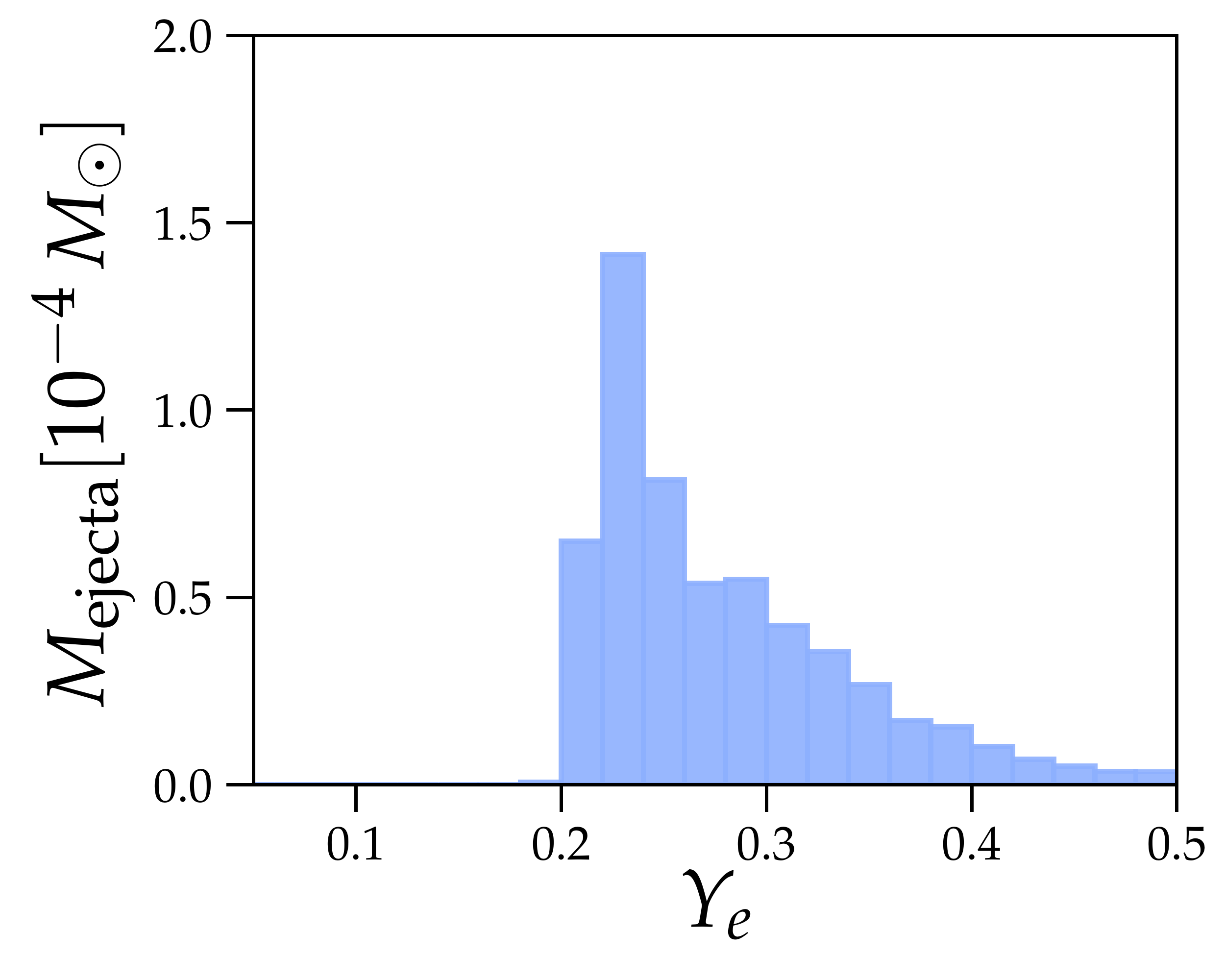}
\end{tabular}
	\caption{Histograms of the radial velocity $v^r$ of the unbound material (top panels), where $r$ is the radius in spherical coordinates, and the $Y_e$ of the unbound material (bottom panels) at different times during the dynamical simulation. Each column corresponds to a different time in the evolution of the system.}.
        \label{fig:vr_hist}
\end{center}
\end{figure*}

In Figure \ref{fig:vr_hist} we present histograms of the radial velocity of the unbound material (top four panels) and its $Y_e$ (bottom four panels) at different times during the evolution of the system. The unbound material is determined via the Bernoulli criterion $-h u_t >1$, where $h$ 
is the relativistic enthalpy of the magnetized fluid. The ejecta show a broad distribution in velocity space. At all times, there exists a significant amount of material with velocities between $0.3c < v^r < 0.4c$. As the system evolves, velocities $0.4c < v^r < 0.48c$ are also seen for a small fraction of the ejecta. The $Y_e$ is a measure of the neutron-richness of matter, given by:
\begin{equation}
    Y_e \equiv \frac{n_p}{n_n+n_p}
\end{equation}
where $n_n$ and $n_p$ are the densities of neutrons and protons respectively. In the ejecta, neutrino-matter interactions drive the $Y_e$ towards higher values i.e. towards less neutron-rich conditions. Most of the ejected material has $Y_e$ values between 0.2 -- 0.3, with a peak around $Y_e$ $\sim$ 0.25. This has interesting implications for $r$-process yields since $Y_e$ values above $\sim$ 0.25 can inhibit the synthesis of any significant amount of heavy elements and the resulting abundance pattern is quite sensitive to the ejecta properties.

\subsection{Tracer Particles and Nucleosynthesis}
\label{sec:methods_nucysn} 

We extract the thermodynamic conditions of the ejected material and the neutrino luminosities it has encountered from the hydrodynamical simulation using Lagrangian tracer particles. The tracers are uniformly spaced to represent regions of constant volume. At the start of the simulation, each tracer particle is assigned a mass that accounts for the density at its location and the volume the particle covers. We place 96000 tracer particles to ensure that a sufficient number of tracer particles are present in the outflow. The particles are advected passively with the fluid and data from the 3D simulation grid are interpolated to the tracer position and recorded as a function of time. The tracers are collected at a surface defined by a chosen radius, $r=$150 M$_{\odot}$ here, and tracer quantities are frozen once the tracer particle crosses this surface.

We track the ejecta composition by post-processing the tracer particles with the open-source nuclear reaction network \texttt{SkyNet} \citep{Lippuner2017}. The network includes 7852 isotopes up to $^{337}$Cn. Forward strong rates are taken from REACLIB \citep{Cyburt2010} and inverse rates are obtained using detailed balance. The weak rates employed here come from \cite{Fuller1982}, \cite{Oda1994}, \cite{Langanke2000}, or from REACLIB. The nuclear masses and partition functions are also obtained from REACLIB. 

The network is started in NSE when the particle temperature drops below 20GK. The network evolves the temperature by calculating source terms due to individual nuclear reactions and neutrino interactions. The neutrino luminosities recorded by the tracers are noisy due to interpolation effects and high time resolution. We perform a moving-window time average of the luminosity data of the form $\nu_{\rm{av},i} = \alpha \cdot \bar{\nu}_{i} + (1 - \alpha) \cdot \nu_{\mathrm{av},i-1}$, where $i$ denotes the current and $i-1$ the previous timestep. The weight function for each data set in the moving average is chosen as $\alpha = 2 (n + 1)^{-1}$, with $n=19$. We keep the luminosities constant after the end of the tracer data.

The dynamical simulation implements a leakage scheme that aims to capture the overall energetics and lepton number exchange due to neutrino emission and absorption. Realistic neutrino luminosities may differ by up to a factor of a few from those extracted by the tracer particles \citep{OConnor2010}. Since the detailed ejecta composition does depend sensitively on the accuracy of the neutrino transport, the uncertainty in our leakage scheme will translate into uncertainties in the $r$-process abundances predicted here. \texttt{SkyNet} allows us to set the neutrino luminosities and energies to constant values during post-processing. We take advantage of this capability to explore the impact of varying the neutrino luminosity over a chosen range i.e. assuming constant values of zero (neglecting neutrinos), $10^{51}$, $10^{52}$ and $10^{53}$ erg s$^{-1}$ (artificially high). The luminosities are set to be the same for electron neutrinos as well as antineutrinos while the average neutrino energies are chosen as 12 and 15 MeV respectively. We compare the resulting abundances to those obtained for the case where we use neutrino luminosities recorded by the tracer particles. The tracer neutrino luminosities are typically a few 10$^{52}$ erg s$^{-1}$ and thus fall between the constant luminosity cases of 10$^{52}$ and 10$^{53}$ erg s$^{-1}$. In this way, we constrain the impact of uncertainties in neutrino luminosities on the final abundances.

\begin{figure}
        \includegraphics[width=0.48\textwidth]{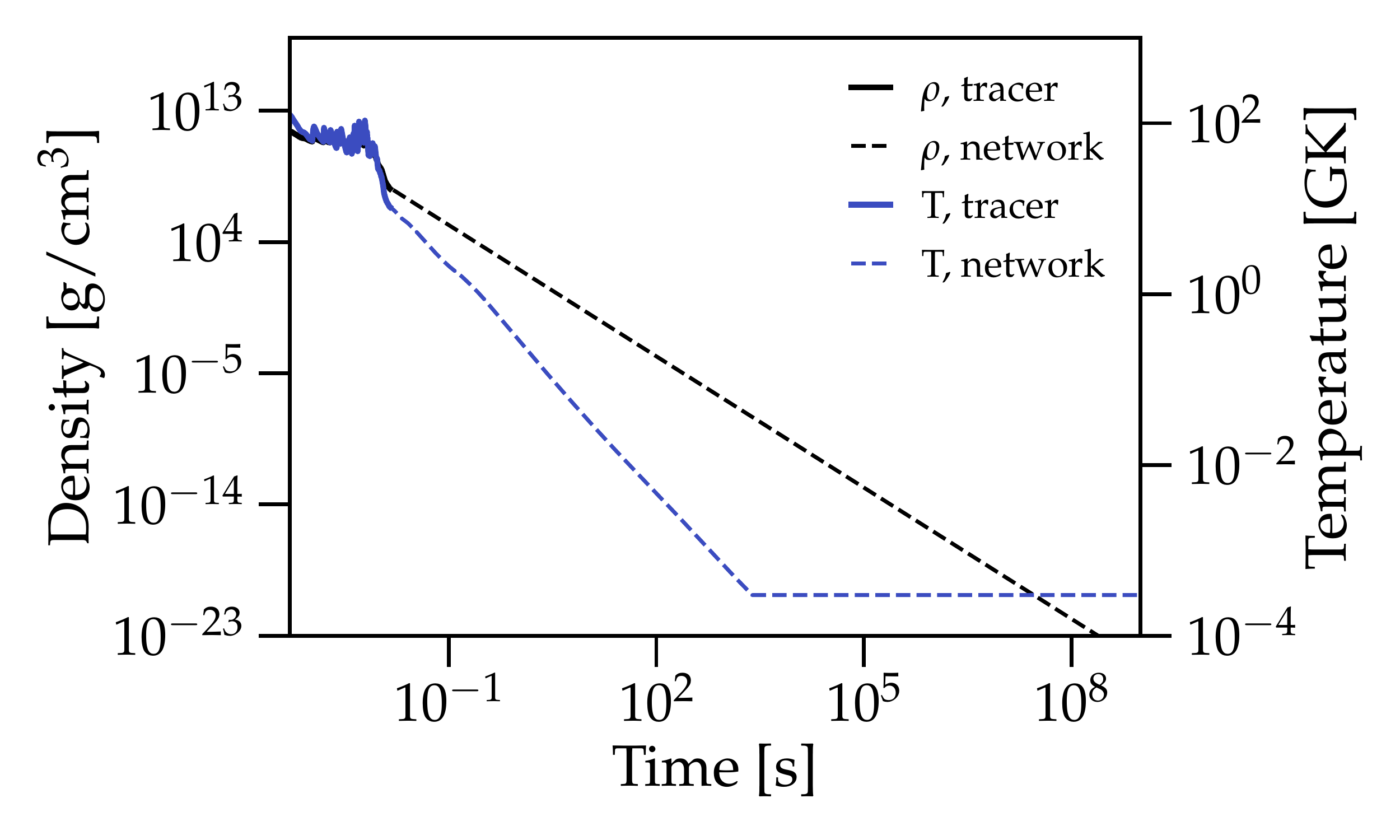}
        \caption{The evolution of density (black lines) and temperature (blue lines) for a representative tracer particle. The solid lines show the original tracer particle data while the dashed lines show the data extrapolated by \texttt{SkyNet} under the assumption of homologous expansion. This evolution corresponds to the case where zero neutrino luminosity is employed during post-processing.}.
        \label{fig:tracer_evol}
\end{figure}

The dynamical simulation, and hence the tracer trajectory, ends within few milliseconds due to the high computational expense of the simulation. However, the requisite conditions for nucleosynthesis usually still exist at this point. The network continues the calculation up to a desired end time by smoothly extrapolating the particle data beyond the end of the trajectory under the assumption of homologous expansion. The network expands the particle using $\rho \propto t^{-3}$ until a minimum temperature is reached. Our calculations are carried out to 10$^9$s, which is sufficient to generate a stable abundance pattern as a function of mass number. The evolution of the tracer temperature and density within \texttt{SkyNet} is shown for a representative tracer particle in Figure \ref{fig:tracer_evol}. 

\subsection{Radiation Transport and Mapping}

\texttt{SNEC} \citep{Morozova2015, Wu2021} is a 1D Lagrangian equilibrium-diffusion radiation hydrodynamics code capable of simulating the hydrodynamical evolution of merger ejecta and the resulting kilonova emission. For kilonova modeling, the nickel heating term relevant for the supernova case in \texttt{SNEC} is replaced by a prescription for radioactive heating due to decay of $r$-process nuclei. 
The time-dependent heating rate is derived in \cite{Wu2021} based on the nucleosynthesis calculations of \cite{Perego2020} by constructing fits over a comprehensive grid of 11700 trajectories, covering a wide range of $Y_e$, entropy $s$ and expansion timescale $\tau$: 0.01$\leq Y_e \leq$0.48, 1.5 $k_\mathrm{B}$ baryon$^{-1}$ $\leq s \leq$ 200 $k_\mathrm{B}$ baryon$^{-1}$ and 0.5 ms $\leq \tau \leq$ 200 ms. At early times $t \lesssim 0.1$ days, fits are constructed using the analytic formula proposed by \cite{Korobkin2012}. At later times $t\gtrsim0.1$ days, a power-law fit is used. The two regimes are joined together via a smoothing procedure and the overall fit describes the heating rate over a time interval ranging from 0.1 seconds to 50 days post-merger. We additionally assume a constant thermalization efficiency of 0.5 taking into account energy loss by neutrinos and partial thermalization of photons, fast electrons and excited nuclei in the expanding ejecta.

To describe the non-trivial photon-matter interactions, we use an effective wavelength-independent (or gray) opacity $\kappa$ to fit the values in \citet{Tanaka2018} computed from detailed energy dependent, radiative transfer simulations. In particular, $\kappa$ is set as a function of the initial $Y_e$ (i.e. before nucleosynthesis) of the ejecta as: 
\begin{equation}
    \kappa = 1+ \frac{9}{1+(4Y_{e})^{12}} \mathrm{cm^2/g}.
\end{equation}
The minimum opacity is 1 cm$^2$g$^{-1}$ and the maximum is 10 cm$^2$g$^{-1}$. The exponent of 12 makes the opacity drop steeply near $Y_e = 0.25$. This spans the overall range of opacities found for low, intermediate and high $Y_e$ material in keeping with the resulting abundances of heavy lanthanides in the ejecta. The steep transition around $Y_e =0.25$ accounts for the steep decrease in typical lanthanide mass fractions from $>$ 0.1 to 10$^{-4}$ that occurs over an extremely narrow range in $Y_e$ centered at $Y_e \approx$ 0.25. 

We do not use the detailed ejecta composition for the kilonova calculation. A simplified version of the Paczynski EOS is employed where the Saha equations are not solved and the correction terms for partial ionization are ignored. We use a value of 2.0 to specify the mean degree of ionization. It is possible to inject additional energy into the ejecta via a thermal bomb but we do not require this capability and therefore set the thermal bomb energy to zero. 

\texttt{SNEC} computes the evolution of the bolometric luminosity as well as AB magnitudes in different bands assuming blackbody radiation. The bolometric luminosity is computed as the sum of the luminosity at the photosphere and the radioactive heating above the photosphere. The location of the photosphere is defined by the optical depth $\tau=2/3$ and the luminosity at the photosphere is given by the usual expression for radiative luminosity calculated at its location:
\begin{equation}
L =-(4\pi r^2)^2 \frac{\lambda a c}{3\kappa} \frac{\partial T^4}{\partial m}    
\end{equation}
where $r$ is the radius, $\lambda$ is the flux-limiter, $a$ is the radiation constant, and $m$ is the mass coordinate. The effective temperature at the photosphere is calculated from the bolometric luminosity $L$ at the photosphere and the photospheric radius $R_{\rm{ph}}$ as $T_{\rm{eff} } = (L / 4 \pi \sigma R_{\rm{ph}}^2)^{1/4}$, where $\sigma$ is the Stefan-Boltzmann constant. To compute the AB magnitudes in different observed wavelength bands, \texttt{SNEC} assumes blackbody radiation at the effective temperature computed at the photosphere and for the layers above the photosphere. While non-thermal radiation is negligible at $T \sim$ 5000 $K$ \citep{Kasen2013}, it becomes important at late times as the ejecta become transparent. It should be noted that the late time light curves are unreliable since the blackbody approach fails as the ejecta become optically thin. 

As input, \texttt{SNEC} requires the radius, temperature, density, velocity, initial $Y_e$, initial entropy and expansion timescale of the outflow as a function of mass coordinate. The entropy $s$ and expansion timescale $\tau$ are used to compute the heating rates and opacities as discussed above. The outflow properties are recorded by measuring the flux of the relevant quantities through a spherical surface at radius $r=$ 100$M_{\odot}$. The unbound material is determined from these 2D data using the Bernoulli criterion and the mass-weighted angle-averaged outflow profile is computed as a function of the enclosed ejecta mass $m$. Since the initial data for \texttt{SNEC} are required at a fixed time, the data are transformed assuming homologous expansion and the radius $r(m)$ is computed from the requirement that $m(r) = 4\pi\int^r_0\rho r^2dr.$

\section{Results}
\label{sec:results}

\subsection{$r$-process Nucleosynthesis}
\label{sec:results_abund}

The $Y_e$ of the ejecta is one of the most critical quantities for determining its ultimate composition. For typical entropy and expansion timescales occurring in the ejecta from BNS mergers, a strong $r$-process occurs for $Y_e \lesssim$ 0.2. The resulting abundance pattern is usually robust and relatively insensitive to the exact $Y_e$, especially for low enough $Y_e$, due to the occurrence of fission cycling. Ejecta with $0.25 \lesssim Y_e \lesssim 0.4$, on the other hand, will not produce substantial abundances of the heavy lanthanides with $A \gtrsim 140$ and the composition is a lot more sensitive to the outflow properties. For $Y_e \gtrsim$ 0.4--0.5, only a weak $r$-process or a composition dominated exclusively by Fe-group nuclei may be obtained. 

\begin{figure}
        \begin{tabular}{c}
        \includegraphics[width=0.45\textwidth]{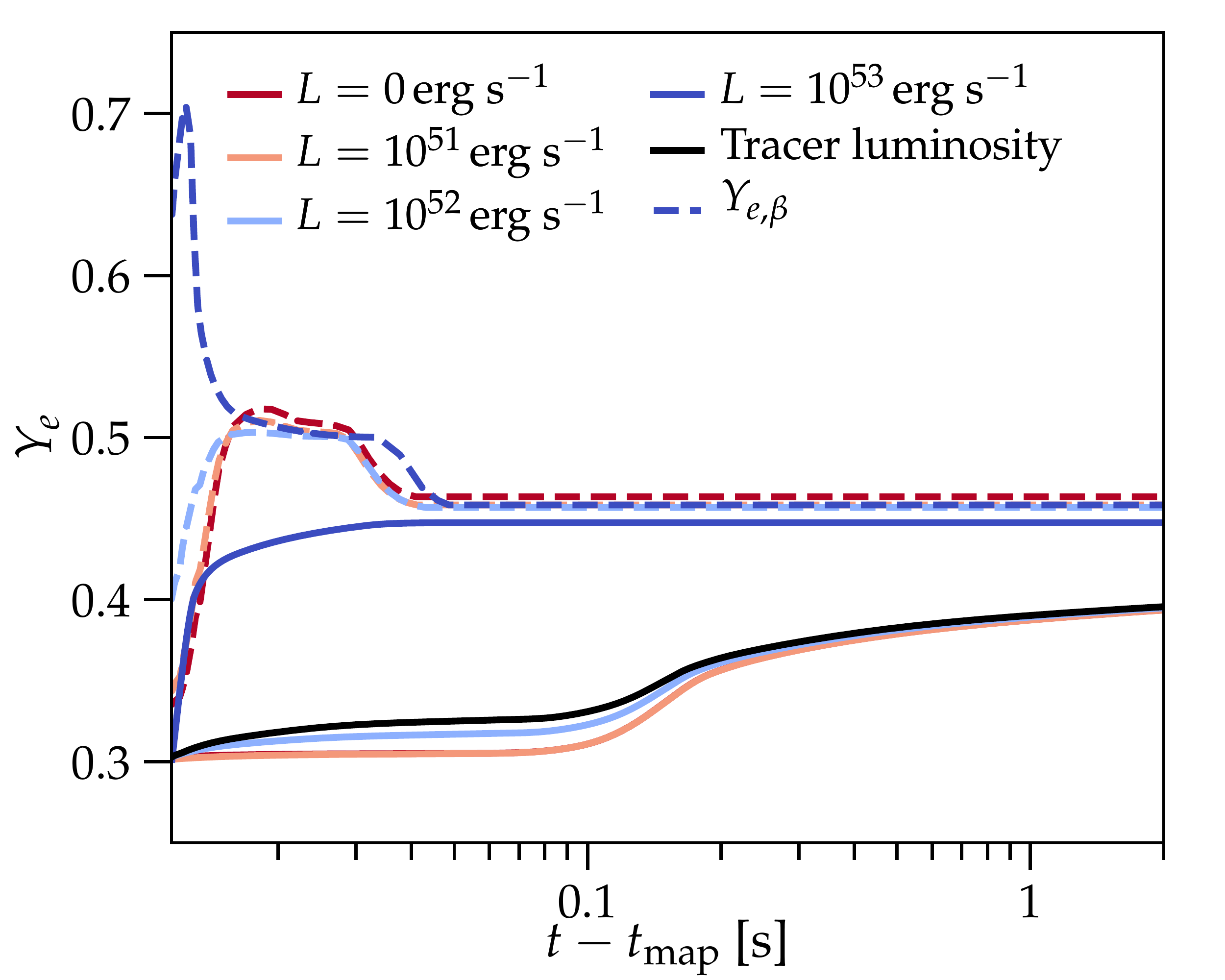}\\
        \includegraphics[width=0.45\textwidth]{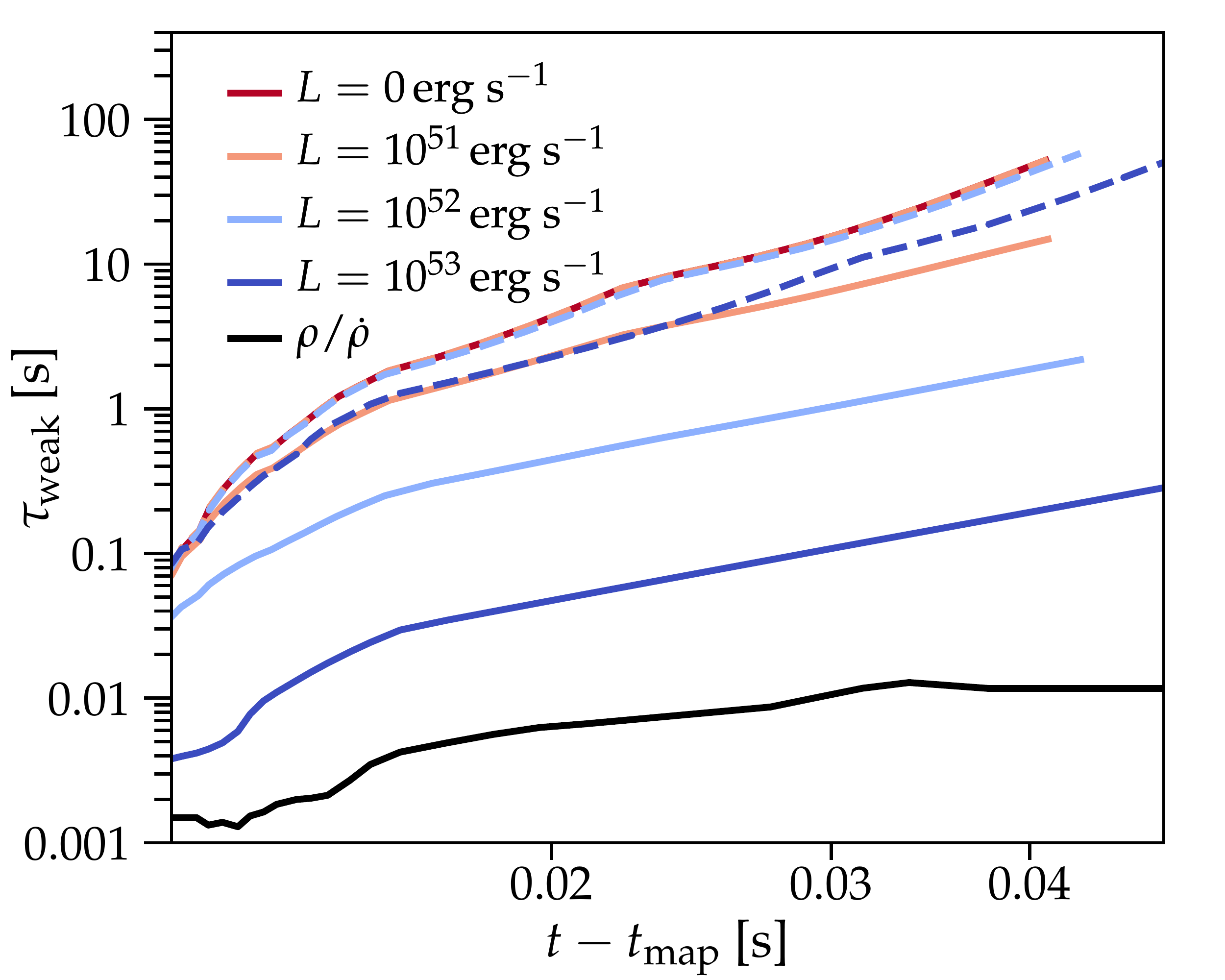}
\end{tabular}
        \caption{The top panel shows the $Y_e$ as a function of time for a representative particle. Different colored lines indicate results for different neutrino luminosities used in the \texttt{SkyNet} calculation and black lines indicate results obtained using the neutrino luminosities recorded by the tracer particles. The dashed lines indicate the evolution of $Y_{e,\beta}$ for the different luminosity settings. The bottom panel shows the weak interaction and dynamical timescales for the same particle. The dashed lines indicate the lepton capture timescale $(\lambda_{e^{-}} +\lambda_{e^{+}})^{-1}$.}
        \label{fig:weak_eq}
\end{figure}

The electron-flavor neutrinos and anti-neutrinos play a crucial role in setting the relative ratio of neutrons to protons and hence the $Y_e$ through the charged-current reactions on free nucleons: 
\begin{equation}
\begin{aligned}
    \nu_{e} + n \rightleftharpoons p + e^- \\
    \bar{\nu}_{e} + p \rightleftharpoons n + e^+.
\end{aligned}
\end{equation}
Weak interactions will drive the $Y_e$ in the neutron-rich wind towards its equilibrium value, given by \citep{Qian1996}:
\begin{equation}
    Y_{e,\beta} \simeq \bigg( 1+ \frac{L_{\bar{\nu}_e}\epsilon_{\bar{\nu}_e} - 2\Delta + 1.2\Delta^2/\epsilon_{\bar{\nu}_e}}{L_{\nu_e}\epsilon_{\nu_e} + 2\Delta + 1.2\Delta^2/\epsilon_{\nu_e}}\bigg)^{-1} \approx 0.4-0.6,
\end{equation}
where $\Delta \equiv (m_n - m_p)c^2$ is the neutron-proton mass difference and $\epsilon_\nu$=$\langle E_{\nu}^2 \rangle/\langle E_{\nu} \rangle$, where $\langle E_{\nu}^n \rangle$ denotes the $n$th neutrino energy moment of the neutrino energy distribution. The range of $Y_{e,\beta}$ comes from the difference between the spectra of electron flavor neutrinos and antineutrinos coming from the HMNS and thus depends on the detailed neutrino transport. Both the neutrino fluxes and the thermodynamic state, e.g. density, of the material also determine the lepton capture rates. As the magnetic field accelerates matter away from the remnant, electron neutrinos will convert the initially neutron-rich composition back towards $Y_e = Y_{e,\beta}$. The final $Y_e$ in the outflow depends on the weak interaction timescale relative to the dynamical timescale but is always driven towards the equilibrium value. This can be seen in Figure \ref{fig:weak_eq}, where we present the $Y_e$ as a function of time for a representative tracer particle along with the corresponding dynamical and weak interaction timescales. Evolution obtained using different luminosity settings during post-processing is shown. The dynamical timescale remains smaller than the weak interaction timescale for all luminosity settings, and the $Y_e$ increases with time but does not attain its weak equilibrium value.

\begin{figure}
        \includegraphics[width=0.47\textwidth]{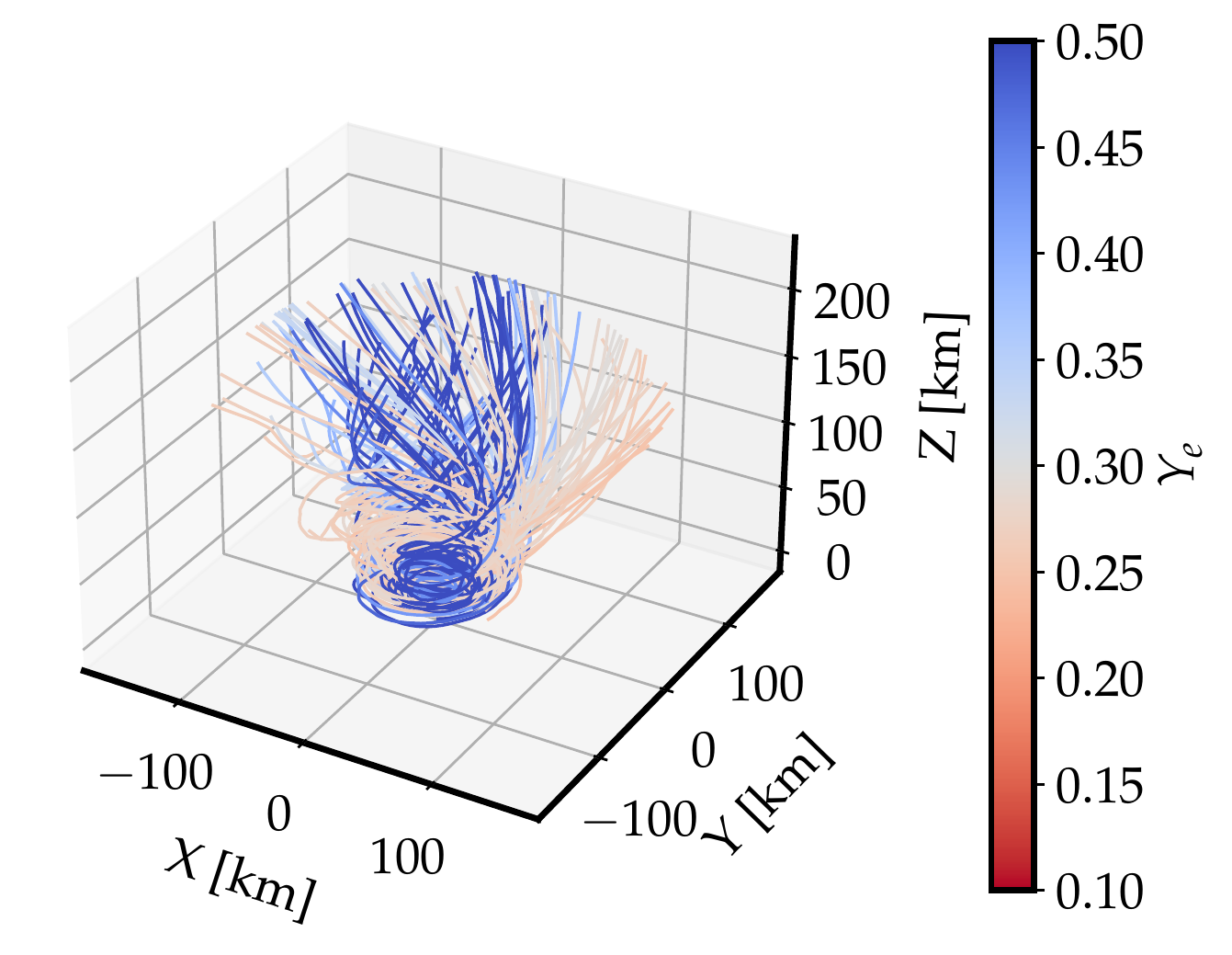}
        \caption{3D spatial trajectories of all tracer particles constituting the HMNS ejecta. The tracers are color-coded to represent the $Y_e$ value attained at the end of the dynamical simulation.}.
        \label{fig:tracer_ye_tracks}
\end{figure}

In general, high neutrino luminosities in the polar region lead to an increase in the $Y_e$ in this region and the longer a particle dwells here, the higher is its final $Y_e$. In Figure \ref{fig:tracer_ye_tracks}, we show the paths of all our tracer particles color-coded by their `final' $Y_e$ (right) at the end of the tracer data. As is evident from the Figure 
, material that is closer to the polar axis has higher $Y_e$ values at the end of the dynamical simulation. This trend in $Y_e$ is preserved during further evolution within \texttt{SkyNet} as the temperature drops below 5~GK and $r-$process nucleosynthesis begins.

\begin{figure}
        \includegraphics[width=0.47\textwidth]{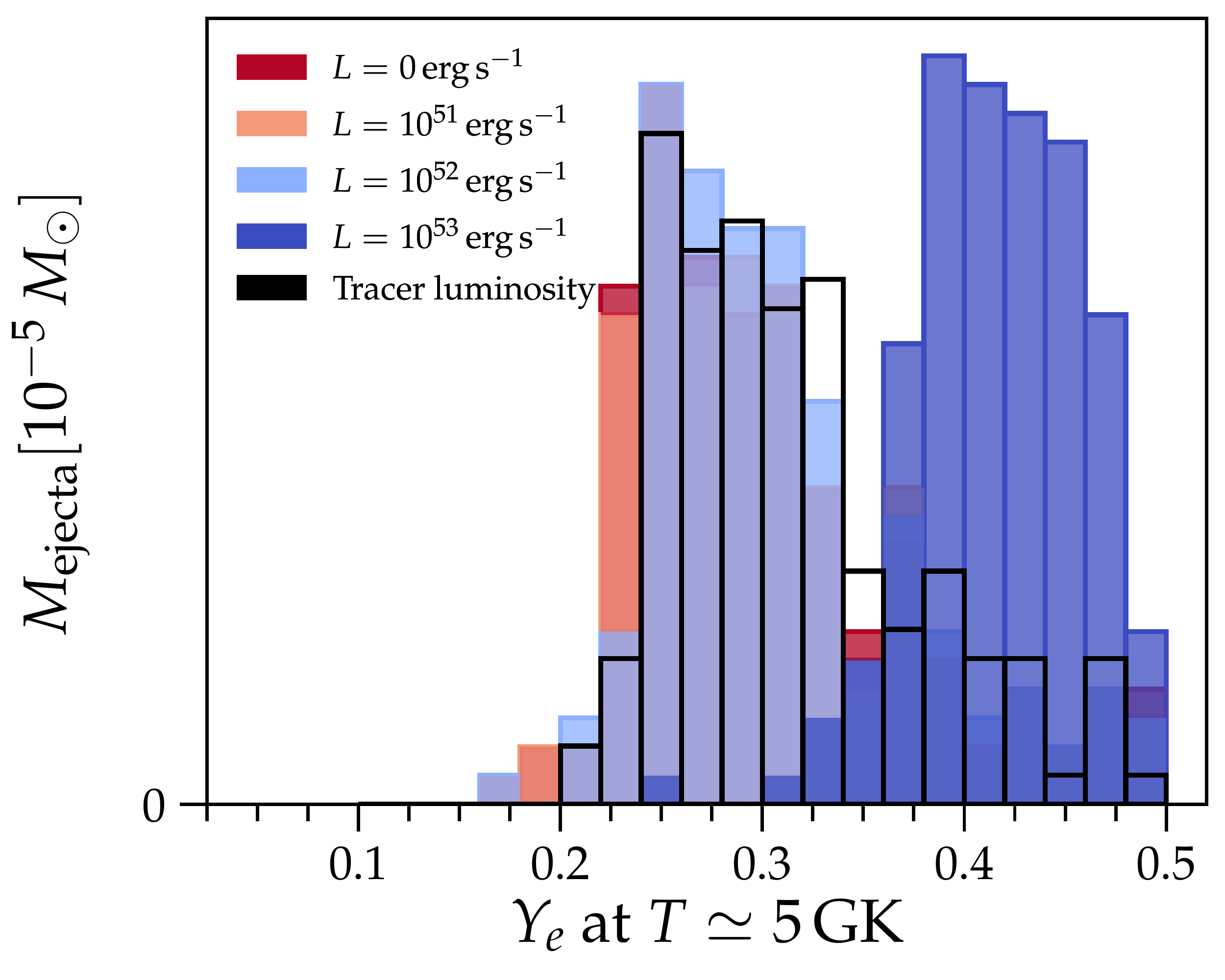}
        \caption{Ejecta $Y_e$ histograms when the particles are last above a temperature of 5GK. The different colors correspond to results obtained with different constant neutrino luminosities used in the network calculations. The black, unfilled histogram shows results obtained using the neutrino luminosities recorded by the tracer particles.}
        \label{fig:ye_hist}
\end{figure}

In Figure \ref{fig:ye_hist},  we show the distribution of the $Y_e$ for all ejected tracers when the temperature of the particles is last above 5~GK, as computed within \texttt{SkyNet}. Since 5~GK is the temperature around which $r$-process nucleosynthesis starts, the $Y_e$ value at this point is the relevant quantity for setting $r$-process yields. We show these quantities for the calculation that uses the leakage neutrino luminosities recorded by the tracer particles, typically of the order of a few times 10$^{52}{\rm erg~s^{-1}}$, as well as for calculations carried out assuming different constant values of the neutrino luminosities. Higher constant neutrino luminosities noticeably shift the peak of the $Y_e$ distribution towards higher values. In the extreme case of $L_\nu = 10^{53}$ erg s$^{-1}$, the ejecta $Y_e$ peaks around $\sim$0.4, above which the synthesis of both the second and third $r$-process peaks is suppressed and only a weak $r$-process can occur.

\begin{figure}
        \includegraphics[width=0.47\textwidth]{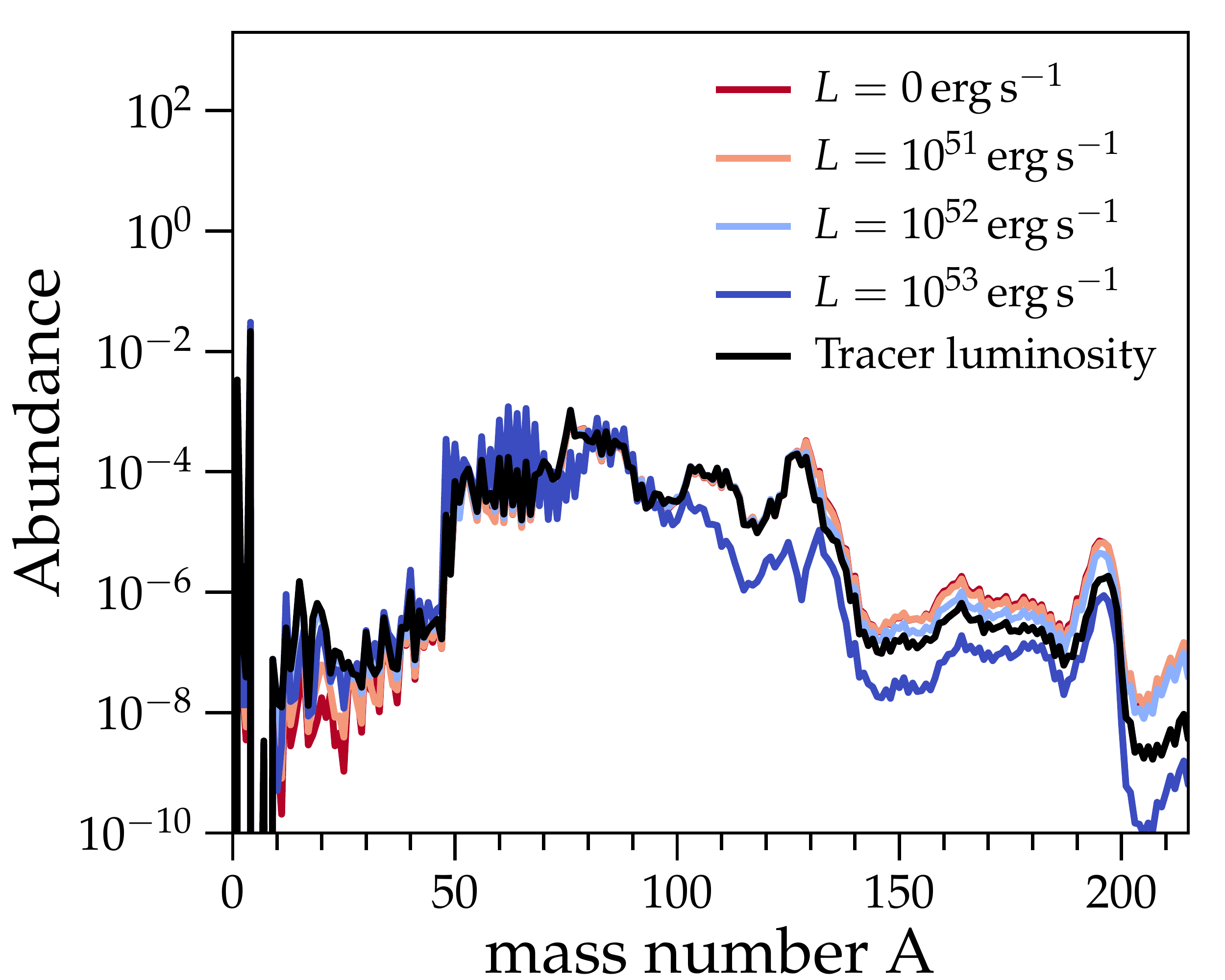}
        \caption{Fractional abundances as a function of mass number $A$ for HMNS ejecta. Different colored lines correspond to results obtained with different neutrino luminosities in the nuclear reaction network calculation.}.
        \label{fig:hmns_abun}
\end{figure}

These variations seen in the $Y_e$ distributions for the different neutrino luminosity scenarios are reflected in the resulting abundance patterns. In Figure \ref{fig:hmns_abun}, we plot the final abundances averaged over all tracers as a function of mass number and compare the results obtained for the four constant luminosity cases and those obtained using the luminosity recorded by the tracer particles. In all cases, there is some production of elements up to the third $r$-process peak but we do not see a robust $r$-process for any of the scenarios. The abundances obtained for the zero luminosity case and the $10^{51}$ erg s$^{-1}$ constant luminosity case are almost identical. Starting with luminosities of $10^{52}$ erg s$^{-1}$, we find that the production of heavy nuclei with $A \gtrsim 140$ is further suppressed. The reduction in the abundances beyond A $\sim$ 140 is accompanied by larger production of nuclei with 50 $\leq$ A $\leq$ 80. However, these three cases still show a significant production of nuclei of the second $r$-process peak. For a constant luminosity of $10^{53}$ erg s$^{-1}$, the production of the second peak is also suppressed and the abundances of the heavy $r$-process elements are reduced by up to a factor of $\sim$10 compared to the zero luminosity case. This is consistent with the expected outcome for an electron-fraction distribution that is peaked around $\sim$0.4.

\begin{figure}
        \includegraphics[width=0.47\textwidth]{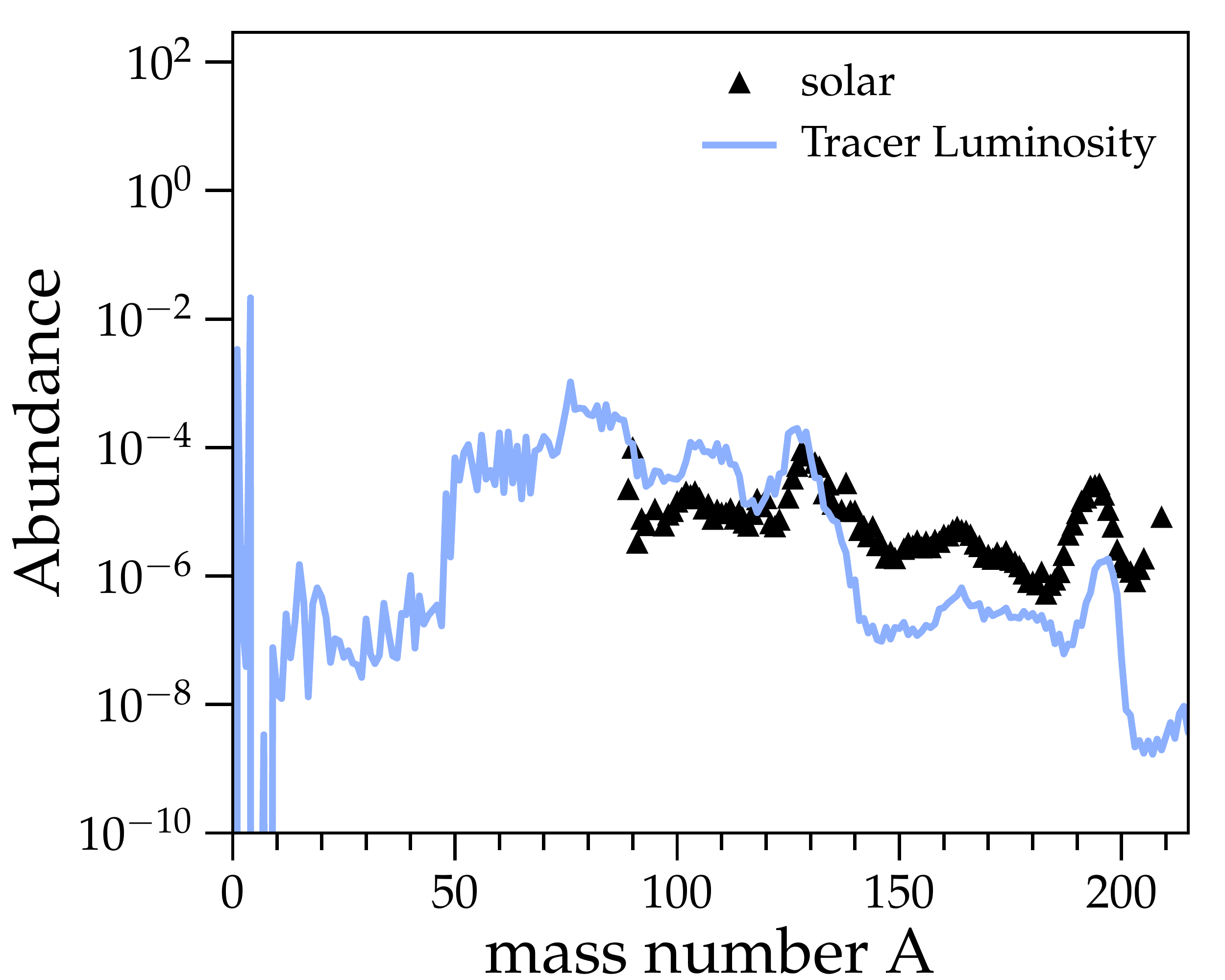}
        \caption{Fractional abundance pattern as a function of mass number $A$ for the HMNS ejecta using neutrino luminosities recorded by the tracer particles during post-processing. Black triangles show the solar abundance pattern scaled to match the second r-process peak at $A$ = 135.}
        \label{fig:compare_sol}
\end{figure}

Finally, in Figure \ref{fig:compare_sol} we present the fractional abundance pattern as a function of mass number obtained for the HMNS ejecta using the tracer luminosities, compared directly to solar abundances. The solar abundance pattern has been scaled to match the second $r$-process peak at $A \sim 135$. We find that the abundances beyond A $\sim$ 140 including the third $r$-process peak are underproduced by up to an order of magnitude or more while lighter nuclei with A $\lesssim$ 135 are overproduced in these ejecta. 

\subsection{Kilonova}
\label{sec:results_kn}

The unstable $r$-process nuclei synthesized in the merger ejecta undergo radioactive decay, heating the ejecta and powering an electromagnetic transient known as a kilonova. The mass and velocity of the ejecta, the radioactive heating rate, the thermalization efficiency of decay products, and the ejecta opacity all play a key role in determining the luminosity evolution of this transient. The opacity in particular sets the time and wavelength(s) at which the ejecta become transparent and in turn depends on the composition of the ejected material. 

\begin{figure}
        \begin{tabular}{c}
        \includegraphics[width=0.45\textwidth]{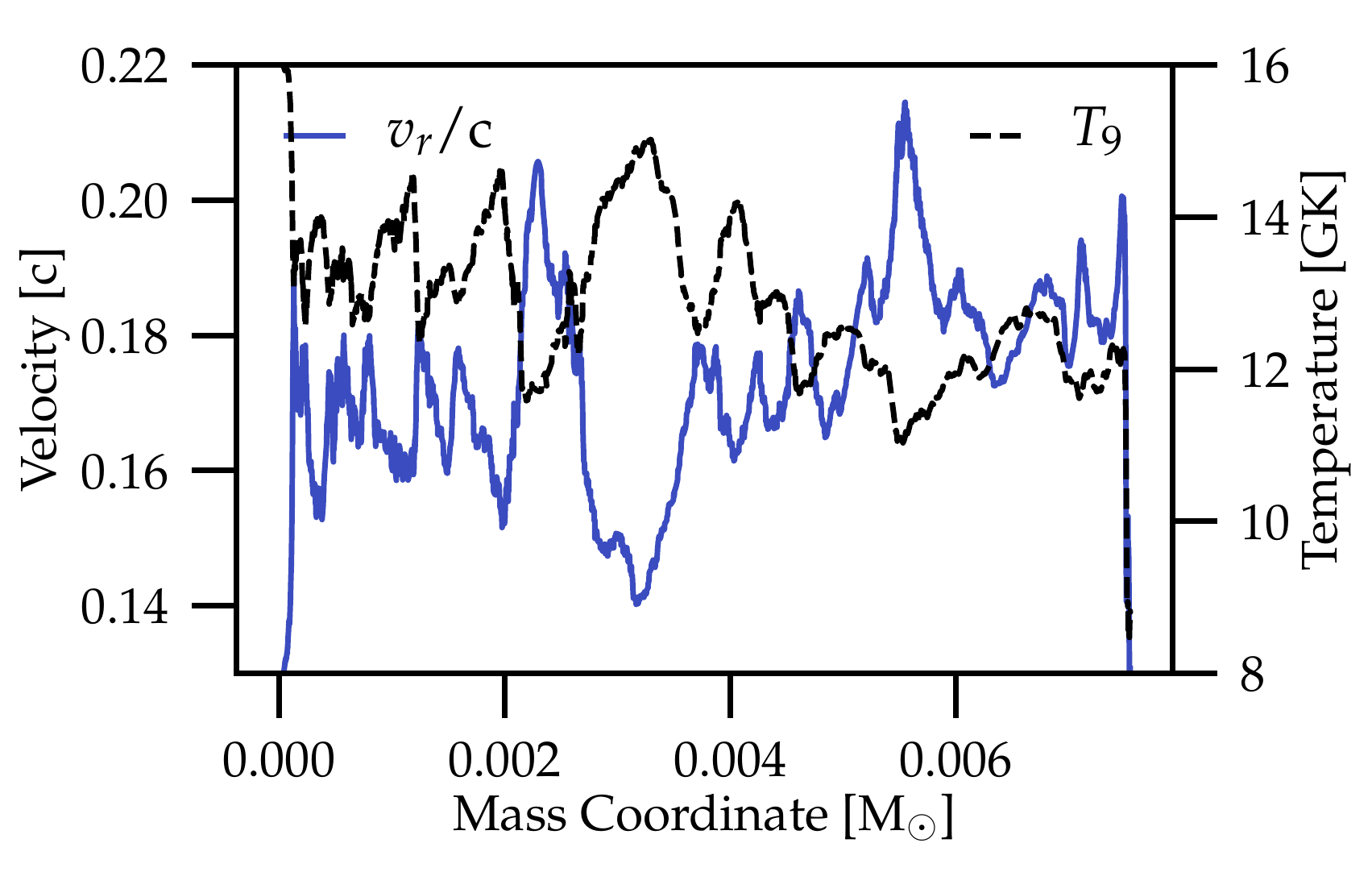}\\
        \includegraphics[width=0.45\textwidth]{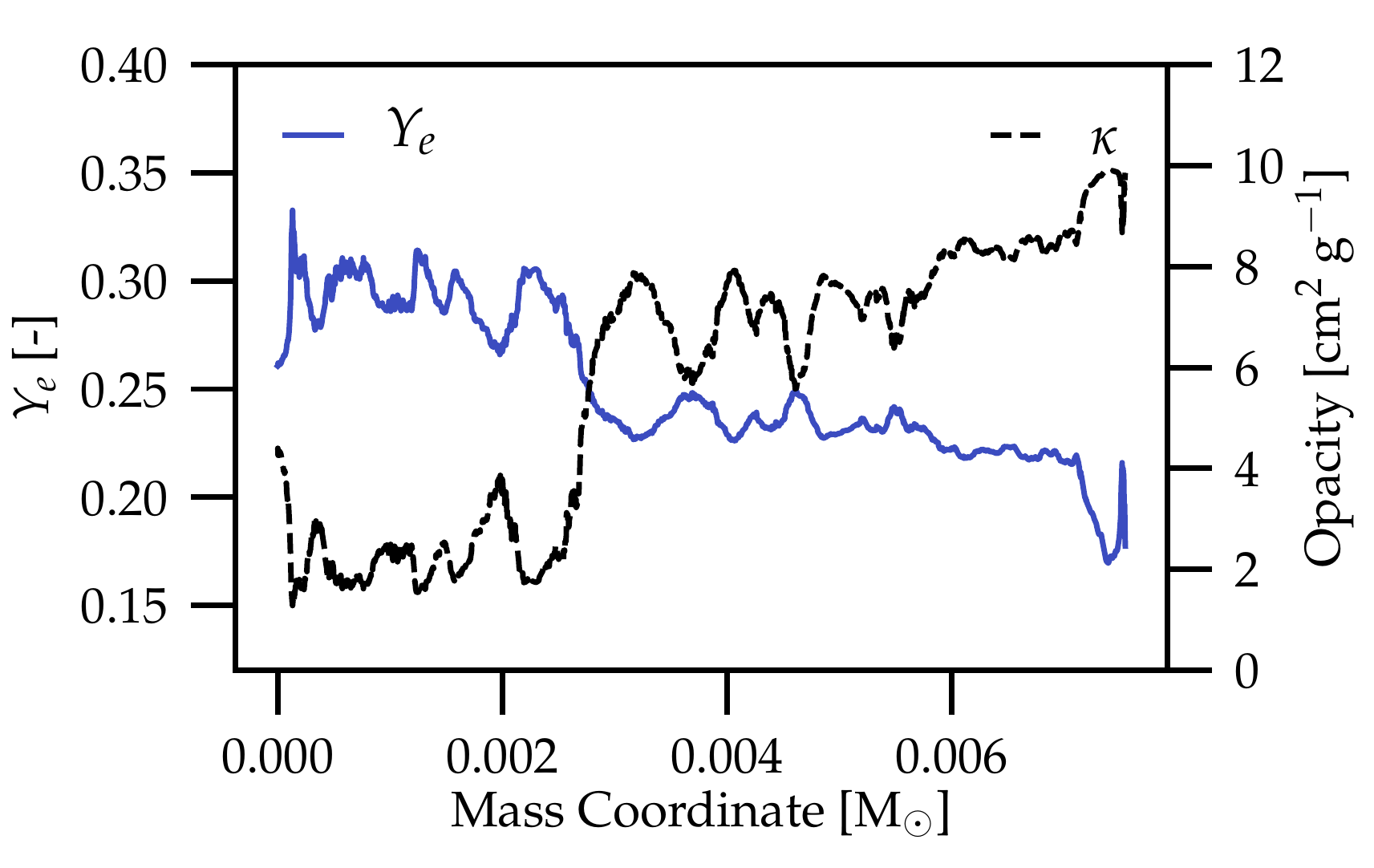}
\end{tabular}
        \caption{Averaged quantities for the HMNS ejecta, as a function of mass coordinate, used as input profiles for \texttt{SNEC}. The top panel shows the velocity (solid blue line) and temperature (dashed black line) of the outflow while the bottom panel shows the $Y_e$ (solid blue line) and the corresponding opacity (dashed black line).}.
        \label{fig:SNEC_input}
\end{figure}

\begin{figure*}
\begin{center}
        \begin{tabular}{ccccc}
        \includegraphics[width=0.2\textwidth]{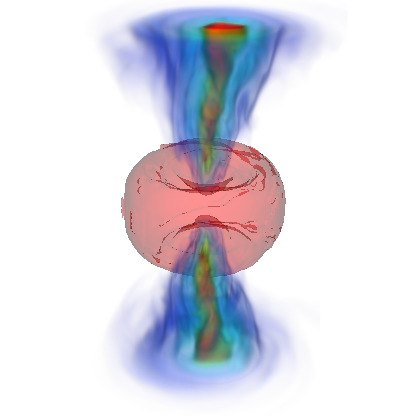}
       \includegraphics[width=0.2\textwidth]{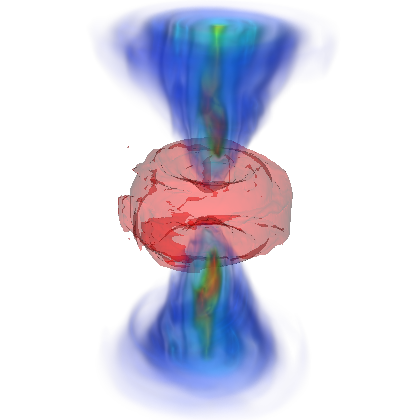} \includegraphics[width=0.2\textwidth]{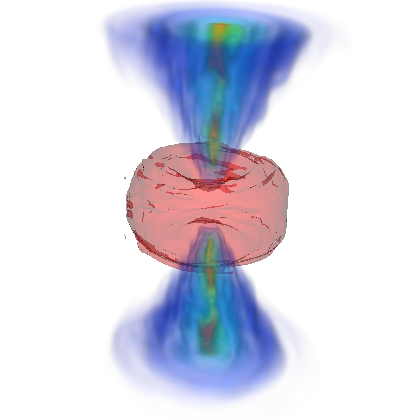}
       \includegraphics[width=0.2\textwidth]{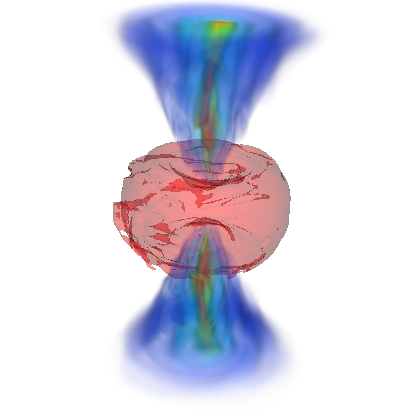} \includegraphics[width=0.2\textwidth]{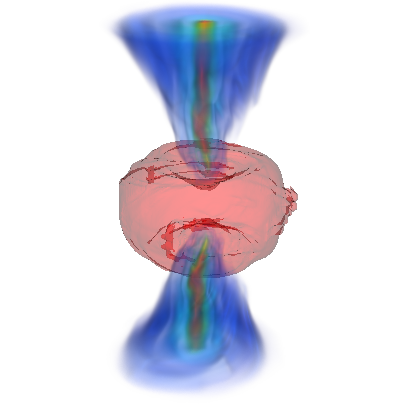}
\end{tabular}
        \caption{Volume renderings of the Bernoulli criterion (blue colormap) indicating unbound material and the disk contour at $\rho = 10^{10}$ g cm$^{-3}$ (red) for model B15-low. The renderings (from left to right) depict the simulation at $t-t_{\rm{map}}$= 9, 12, 15, 18 and 21 ms. The $z$-axis is the rotation axis of the HMNS and we show the innermost 357 km. The colormap is chosen such that blue corresponds to material with lower Lorentz factors $-h u_t \simeq$ 1, while yellow corresponds to material with $-h u_t \simeq$ 1.5, and red to material with $-h u_t \simeq$ 2$-$5. We note that for rendering purposes we have excluded part of the unbound ejecta in the equatorial region.}
      \label{fig:3d_renders}
\end{center}
\end{figure*}

In Figure \ref{fig:SNEC_input}, we present the averaged velocity, temperature, electron fraction and opacity profiles of the HMNS outflow used as input for \texttt{SNEC}. The total ejecta mass is $\sim 7.5 \times 10^{-3} M_{\odot}$, most of it extremely hot with temperatures between $\sim$ 9--16 GK and moving at velocities between $\sim$ 0.15 -- 0.2$c$. In Figure \ref{fig:3d_renders}, we show 3D volume renderings of the Bernoulli criterion for the outflow, along with an isocontour plot for a density of 10$^{10}$ g cm$^{-3}$. The narrow red funnel aligned with rotation axis (z-axis) shows the mildly relativistic jet while blue corresponds to material with lower Lorentz factors. The changing behavior of ejecta velocity as a function of mass coordinate broadly aligns with the time evolution seen in these 3D renderings. In particular, there is a brief decrease in the ejecta velocity over time before the outflow resumes steady-state operation, which is reflected in the dip in the velocity profile around mass coordinate $\sim 3.0 \times 10^{-3} M_{\odot}$. The electron fraction of the ejecta increases systematically as we move inward in mass coordinate. A negligibly small amount of material, located in the outermost layers, has $Y_e \lesssim$ 0.2 while a substantial fraction of the ejecta with a total mass of $\sim 4.8 \times 10^{-3}$ $M_{\odot}$ has slightly higher values of $0.2 \lesssim Y_e \lesssim 0.25$. Finally, material below mass coordinate $\sim 2.7 \times 10^{-3}  M_{\odot}$ has intermediate values of $ 0.25 \lesssim Y_e \lesssim 0.35$. The corresponding opacities are $\sim$10 cm$^2$ g$^{-1}$ for the outermost, low-$Y_e$ layers of the ejecta, between $\sim 5-10$ cm$^2$ g$^{-1}$ for the bulk of the ejecta, declining as we move inward  in mass coordinate to $\sim 1-5.5$ cm$^2$ g$^{-1}$ for the inner ejecta with $Y_e \gtrsim$ 0.25. 

\begin{figure}
    \includegraphics[width=0.49\textwidth]{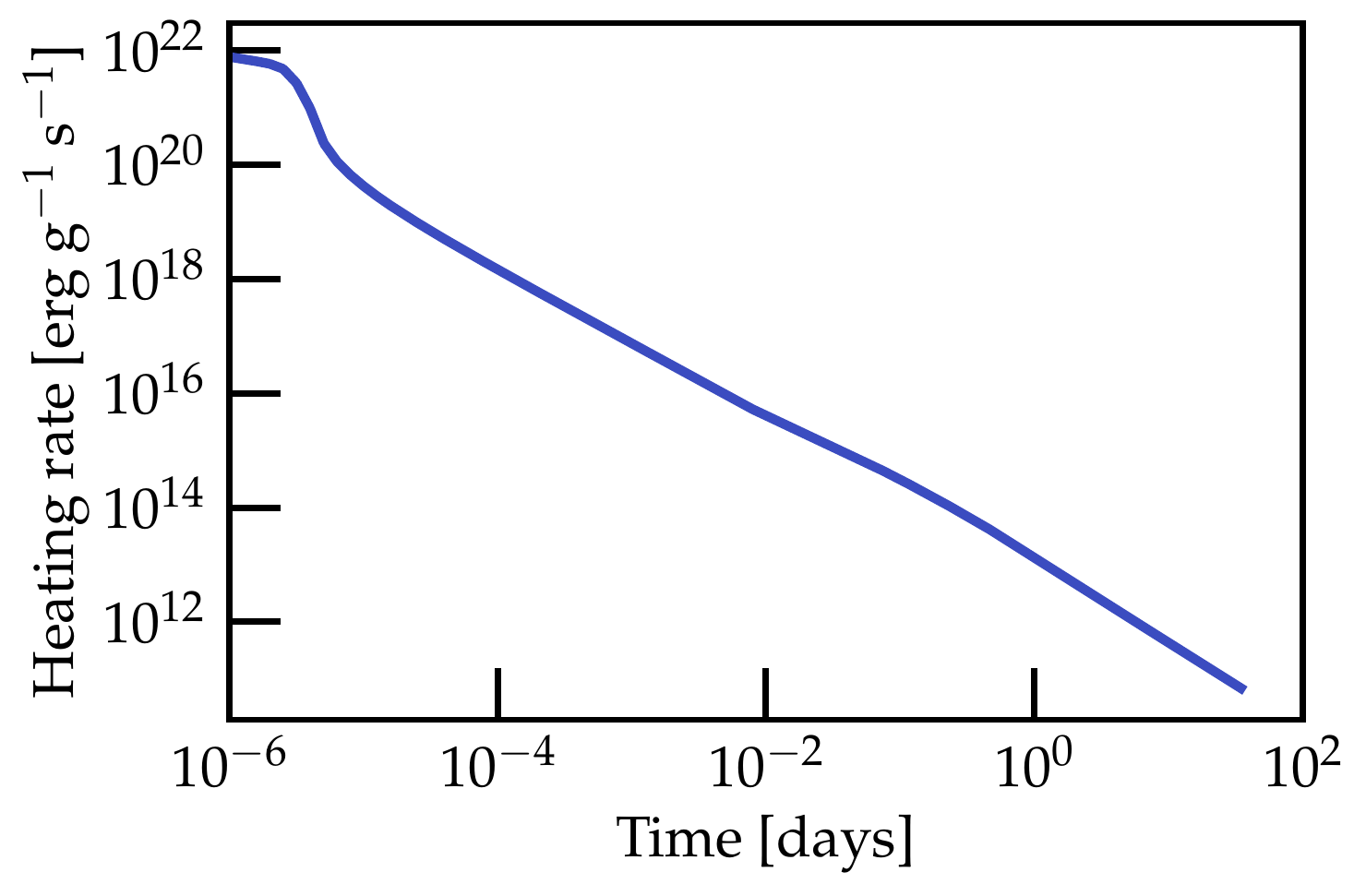}
        \caption{Radioactive heating rate per unit mass due to the decay of $r$-process material as a function of time.}.
      \label{fig:kn_heat}
\end{figure}

\begin{figure*}
\begin{center}
\includegraphics[width=0.98\textwidth]{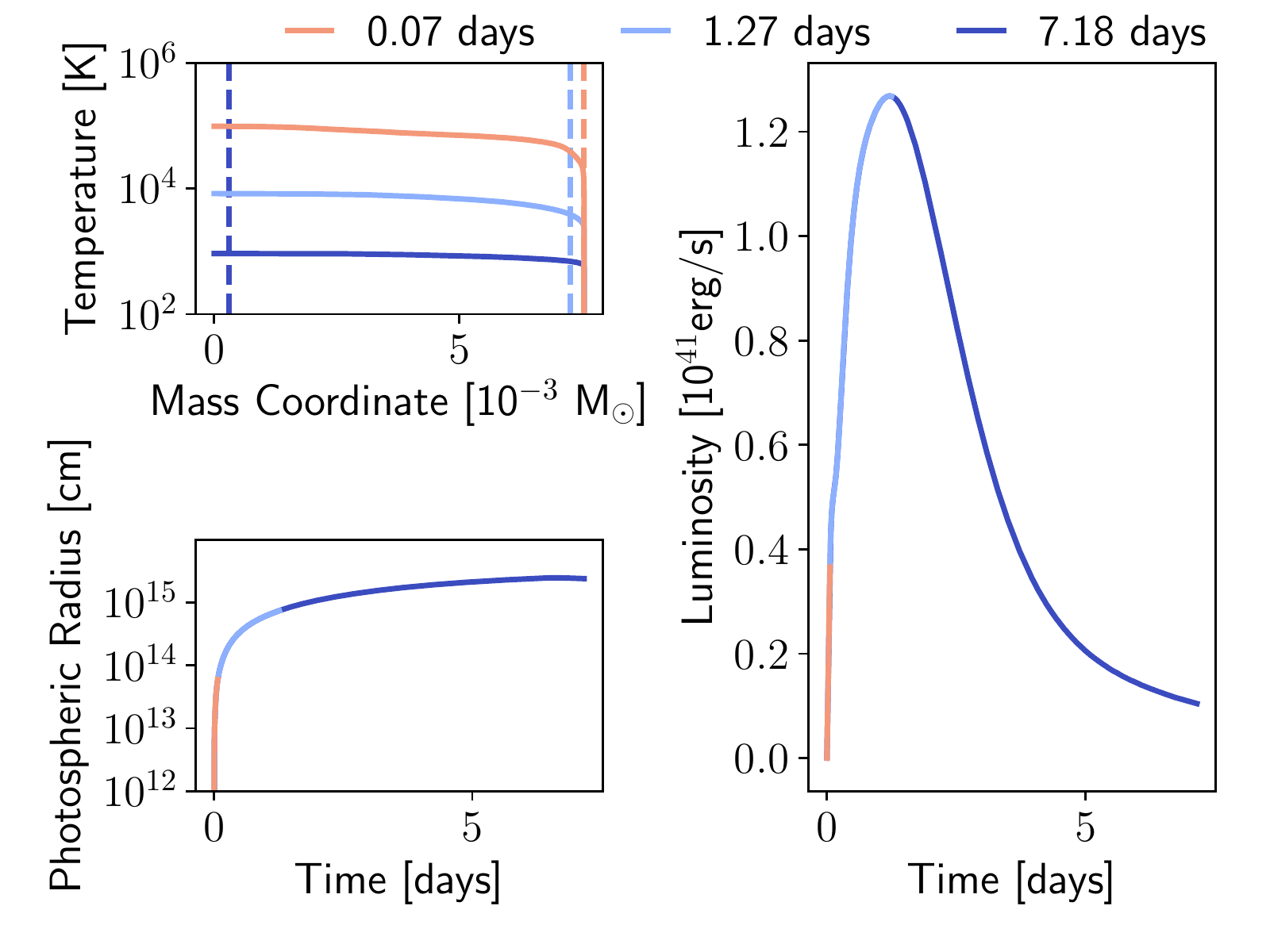}
\caption{Evolution of the temperature profile of the outflow, the radius of the photosphere, and the bolometric luminosity of the kilonova as computed with \texttt{SNEC}. The three colors correspond to three different times: the peach line depicts the evolution up to a time of 0.07 days, the light blue line up to 1.27 days and the deep blue line up to 7.18 days. The vertical dashed lines in the top left panel represent the location of the photosphere in mass coordinate at these three times.}
\label{fig:SNEC_evol}
\end{center}
\end{figure*}

The majority of the energy release due to radioactive decay of $r$-process nuclei happens on a timescale of $\sim$ seconds, as can also be seen in Figure \ref{fig:kn_heat}. However, most of the initial heat in the ejecta is lost through adiabatic expansion because the optical depth is very high and the thermal energy cannot yet escape as radiation. 
In Figure \ref{fig:SNEC_evol}, we show the evolution of the temperature of the outflow in \texttt{SNEC}, along with the photospheric radius and bolometric luminosity. The location of the photosphere is shown by the vertical dashed lines. From Figure \ref{fig:SNEC_evol}, we can see that the ejecta have cooled through expansion from initial temperatures of the order of $\sim$10$^{10}$ K to temperatures around $\sim$ 10$^{5}$ K within a couple of hours. As the ejecta expand and the density decreases, the photon diffusion time also decreases. The photosphere moves inward in mass coordinate (although still outward in radius) through the relatively high-opacity material in the outer layers. Significant electromagnetic emission becomes possible when the density is sufficiently low such that the photons can escape the ejecta on the timescale of expansion $\sim R/v$. This condition sets the characteristic radius at which the luminosity peaks and the corresponding time to peak. For these ejecta, the bolometric luminosity hits its peak at around one day, as expected for a low-mass, high-velocity outflow. The value of the peak luminosity depends sensitively on opacities and the amount of radioactive heating that occurs over the time to peak. The ejecta continue to expand and cool and the photosphere reaches the center at $\sim$ 8.19 days when the entire ejecta become transparent to radiation. Beyond this point, the concept of a photosphere is no longer relevant and the observed luminosity is governed entirely by the radioactive heating rate. 

\begin{figure}
    \includegraphics[width=0.49\textwidth]{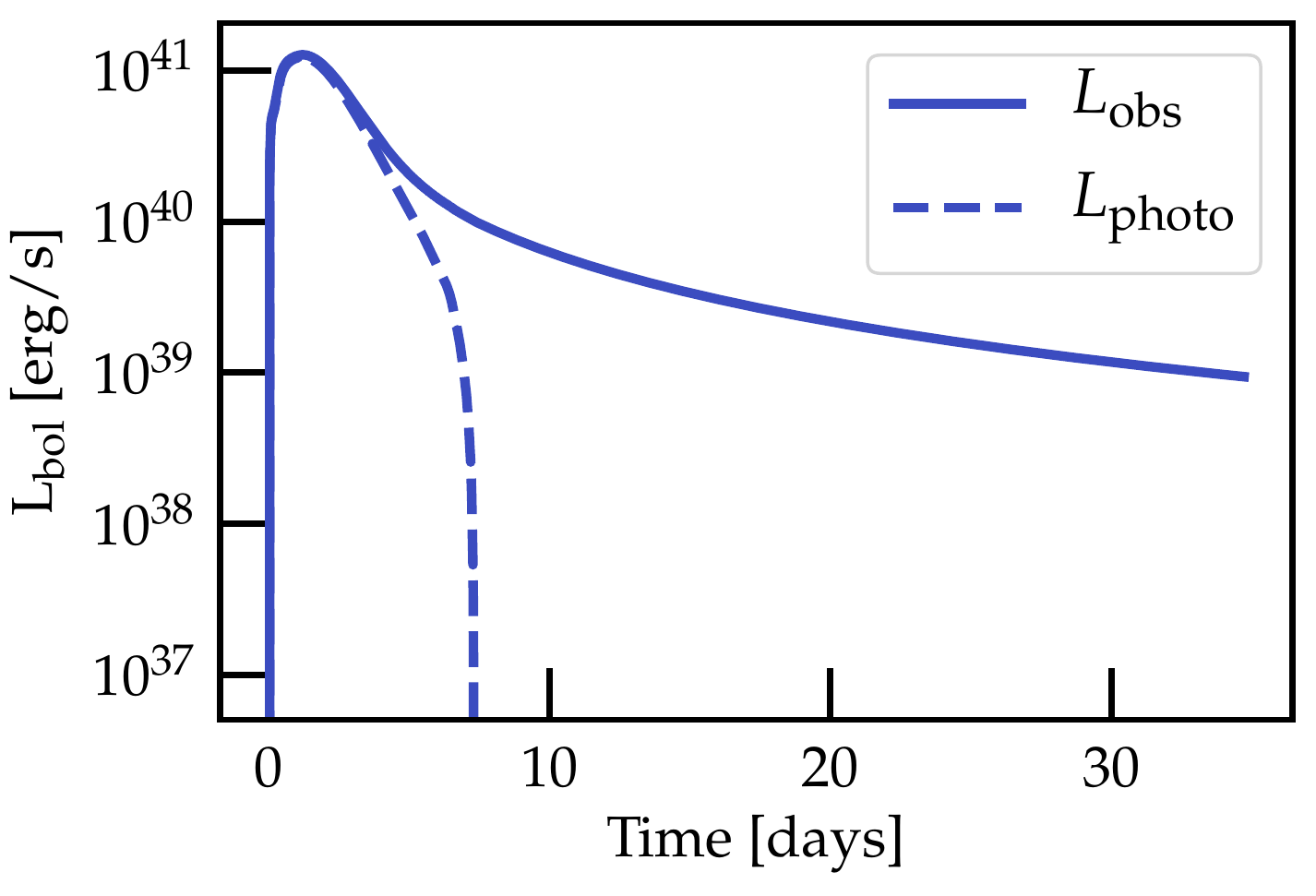}
	\caption{ Observed bolometric luminosity (solid line) and the luminosity at the photosphere (dashed line) as a function of time.}.      \label{fig:kn_lum}
\end{figure}

The bolometric light curve of the kilonova is shown in Figure \ref{fig:kn_lum}, along with the luminosity at the photosphere. Given the velocity and opacity distribution of our ejecta, a simple spherical free-expansion model would suggest a peak timescale of $\sim$ 1 to 3 days. The peak luminosity, which depends on the amount of radioactive heating that occurs on the peak timescale, can be estimated to be of the order of $\sim$ 10$^{41}$ erg s$^{-1}$. In keeping with these estimates, the bolometric light curve peaks at $L_{\rm{obs}}\sim$1.27 $\times 10^{41}$ erg/s at roughly 1.2 days. The effective temperature at the photosphere is $\sim$4257 K at peak luminosity.

\begin{figure}
    \begin{tabular}{c} 
    \includegraphics[width=0.49\textwidth]{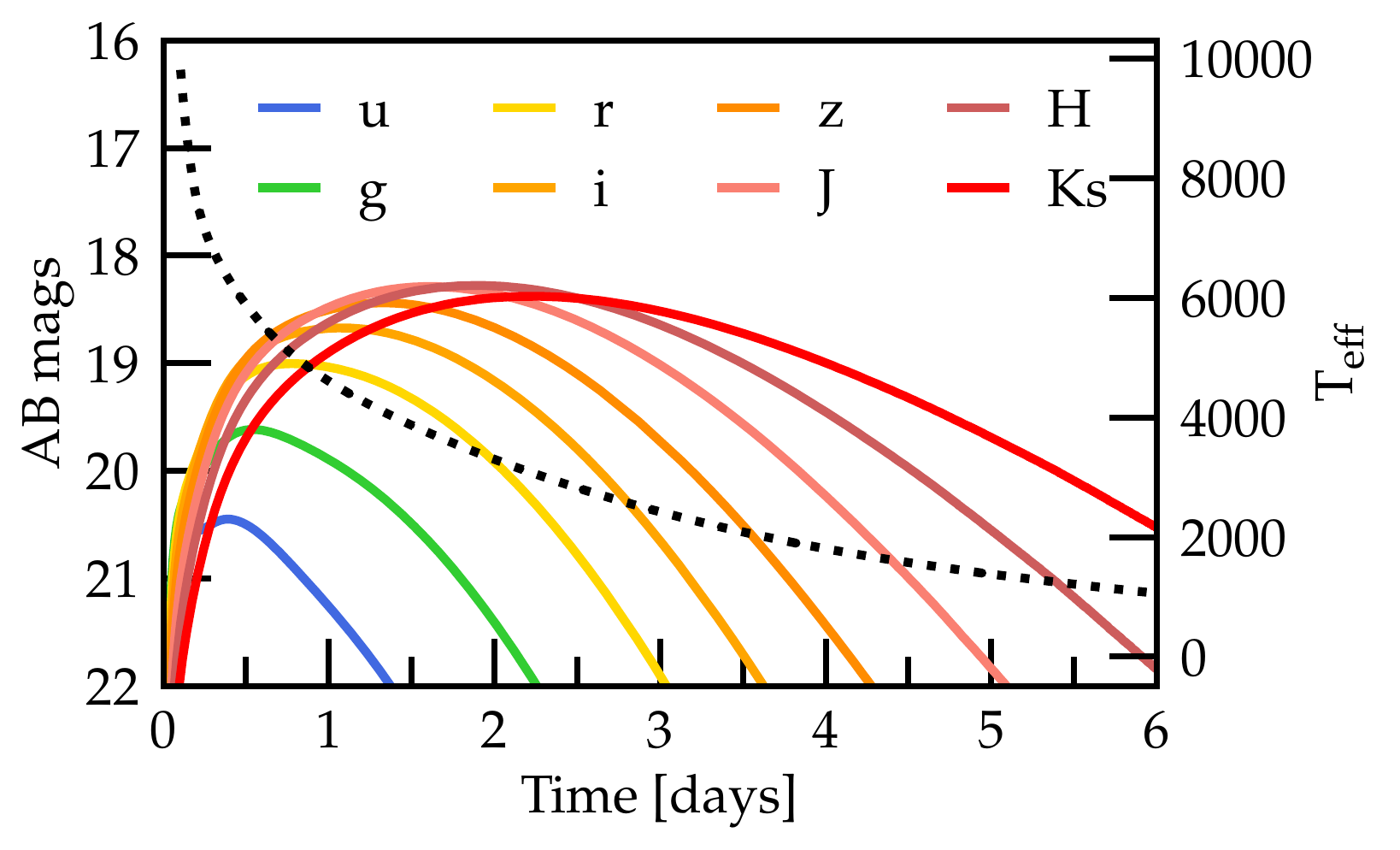}
\end{tabular}
        \caption{AB magnitudes of the kilonova in the $ugriz$JHK$_s$ bands are shown by solid colored lines. The dotted black line gives the evolution of the effective temperature at the photosphere. The distance between the observer and the kilonova is taken to be 40 Mpc, same as the approximate distance to AT2017gfo.
        }.
        \label{fig:kn_mag}
\end{figure}

In Figure \ref{fig:kn_mag}, we present the corresponding AB magnitudes in optical ($ugriz$) and near-infrared (JHK$_s$) filters, produced under the assumption of blackbody emission at the photosphere and for the layers above the photosphere. The distance between the observer and the kilonova is taken to be 40 Mpc, same as the approximate distance to AT2017gfo. The black dotted line gives the evolution of the effective temperature computed at the photosphere. The shorter wavelength bands peak first since the effective temperature at the photosphere decreases as the ejecta expand and cool. By the time the bolometric luminosity hits its peak, the effective temperature has dropped to $\sim$4257 K and continues to decrease rapidly, shifting the spectral energy distribution towards longer wavelengths and further into the infrared. The lowest peak magnitudes and hence the brightest emission is seen in the J and H bands around day 2, both attaining a magnitude of $\sim$18.3 at 1.6 and 1.9 days respectively. Although the fast evolution of the bolometric luminosity and a peak timescale of around a day are consistent with low-opacity high-$Y_e$ outflows that produce a kilonova with a blue spectral peak, the band light curves for these ejecta peak in the infrared. This is due to the presence of high opacity material in the outer layers of the ejecta. By the time the ejecta expand enough to radiate efficiently, they have also cooled down substantially and the majority of the observed emission happens at longer wavelengths.  

\section{Summary and Discussion}
\label{sec:summary}

We have predicted $r$-process abundances and kilonova emission for outflows from HMNS remnants based on a dynamical 3D GRMHD simulation. The simulation includes a nuclear EOS and neutrino effects through a leakage scheme. To bracket possible uncertainties in the composition due to the approximate neutrino treatment, we have employed a range of constant neutrino luminosities during nucleosynthetic post-processing. We have mapped the outflow to an equilibrium-diffusion radiation hydrodynamics code to predict the bolometric light curve of the resulting kilonova as well as AB magnitudes in $ugriz$JHK$_s$ bands assuming blackbody emission. Our main findings are:
\begin{itemize}
    \item The ejecta show a wide distribution in their $Y_e$, peaking in the $\sim$0.25--0.4 range depending on the choice of neutrino luminosity during post-processing. We do not find a robust third $r$-process peak and abundances beyond the second peak are reduced for constant luminosities above 10$^{52}$ erg s$^{-1}$. This change in abundances aligns with the shifting peak of the $Y_e$ distribution towards higher values.
    \item The averaged spherically-symmetric profiles of the ejecta shows outflow velocities between $\sim$ 0.15 -- 0.2$c$ and $Y_e$ between $\sim$ 0.2 -- 0.35, with a total ejecta mass of $\sim 7.5 \times 10^{-3}$ $M_{\odot}$. The low $Y_e$, high-opacity ejecta lies ahead of the high $Y_e$, low-opacity ejecta in mass coordinate. 
    \item The bolometric light curve of the kilonova peaks at roughly one day with a luminosity of $\sim$ 10$^{41}$ erg s$^{-1}$ at peak, as expected for a low mass and high velocity outflow and given the heating supplied by decay of $r$-process nuclei. 
    \item The brightest emission is seen in the J and H infrared bands with a peak magnitude of $\sim$18.3, corresponding to effective photospheric temperatures of the order of $\sim$4000K. 
\end{itemize} 

These HMNS ejecta represent a distinct and important part of BNS outflows in addition to the dynamical ejecta and accretion disk winds. For longer-lived remnants they could constitute the dominant component setting the $r$-process yields and kilonova properties. The bulk of the dynamical ejecta have much lower $Y_e$ in comparison and a comparable total mass, while the disk winds have much lower velocities than HMNS ejecta. Both dynamical ejecta and disk winds are expected to make substantial amounts of lanthanides and produce a redder kilonova. For magnetically-accelerated HMNS winds, however, the ejecta $Y_e$ depends sensitively on the dynamics and it is possible for the $Y_e$ in the outflow to be high enough to inhibit the synthesis of a substantial mass-fraction of lanthanides, and for the kilonova to consequently peak at visible wavelengths. This work is the first to predict the composition of HMNS ejecta and the nature of the resulting kilonova using realistic ejecta properties extracted from a 3D GRMHD dynamical simulation. We have shown that HMNS outflows will not produce a robust third $r$-process peak and the lanthanide fraction in these ejecta depends on the neutrino luminosity encountered. Additionally, given the ejecta properties obtained here, the kilonova observed will peak around one day and in the infrared bands.

However, this picture may change for longer-lived remnants and if the evolution of the ejecta properties is followed over a longer time period. For longer remnant lifetimes, the total ejecta mass will be higher, resulting in a more luminous kilonova.  We time-extrapolated the ejecta profiles assuming a HMNS lifetime of 100ms to obtain a larger total ejecta mass $\sim10^{-2} M_{\odot}$ and found that the resulting kilonova is more luminous but the behavior across different wavelength bands remains mostly unchanged. The electron fraction distribution of the outflow will also shift to higher values over time. While the nucleosynthesis predictions presented in this work account for the effects of uncertain neutrino luminosities and the $Y_e$ evolution of the tracer particles beyond the end of the hydrodynamical data, these aspects are currently not accounted for in the kilonova calculation. We directly use the outflow properties recorded at an extraction radius of $r=100 M_{\odot}$ to predict the resulting kilonova emission. Especially for neutrino luminosities $\sim$10$^{53}$ erg/s, it may be possible to remove the high opacity material or lanthanide curtain \citep{Kasen2015, Wollaeger2018, Nativi2021} in the outer layers of the ejecta, allowing the kilonova to peak in bluer bands.  Additionally, we currently neglect the ejecta composition and compute band light curves assuming blackbody radiation at the effective temperature instead of using detailed opacities.

Further exploration of the kilonova counterpart will require tracking the outflow properties over a longer time period than we have considered here, taking into account the ejecta morphology as well as its composition, and employing detailed wavelength-dependent opacities. The total mass, velocity and electron-fraction distribution of these outflows broadly align with the derived values for the blue kilonova component associated with GW170817, indicating that detailed long-term end-to-end modeling is needed to definitively answer the question of whether HMNS ejecta can produce a blue kilonova.

\section*{Acknowledgements}

DR acknowledges funding from the U.S. Department of Energy, Office of
Science, Division of Nuclear Physics under Award Number(s)
DE-SC0021177 and from the National Science Foundation under Grants No. 
PHY-2011725, PHY-2020275, PHY-2116686, and AST-2108467.





\bibliographystyle{mnras}
\bibliography{main} 

\begin{thebibliography}{}
\makeatletter
\relax
\def\mn@urlcharsother{\let\do\@makeother \do\$\do\&\do\#\do\^\do\_\do\%\do\~}
\def\mn@doi{\begingroup\mn@urlcharsother \@ifnextchar [ {\mn@doi@}
  {\mn@doi@[]}}
\def\mn@doi@[#1]#2{\def\@tempa{#1}\ifx\@tempa\@empty \href
  {http://dx.doi.org/#2} {doi:#2}\else \href {http://dx.doi.org/#2} {#1}\fi
  \endgroup}
\def\mn@eprint#1#2{\mn@eprint@#1:#2::\@nil}
\def\mn@eprint@arXiv#1{\href {http://arxiv.org/abs/#1} {{\tt arXiv:#1}}}
\def\mn@eprint@dblp#1{\href {http://dblp.uni-trier.de/rec/bibtex/#1.xml}
  {dblp:#1}}
\def\mn@eprint@#1:#2:#3:#4\@nil{\def\@tempa {#1}\def\@tempb {#2}\def\@tempc
  {#3}\ifx \@tempc \@empty \let \@tempc \@tempb \let \@tempb \@tempa \fi \ifx
  \@tempb \@empty \def\@tempb {arXiv}\fi \@ifundefined
  {mn@eprint@\@tempb}{\@tempb:\@tempc}{\expandafter \expandafter \csname
  mn@eprint@\@tempb\endcsname \expandafter{\@tempc}}}

\bibitem[\protect\citeauthoryear{{Abbott} et~al.,}{{Abbott}
  et~al.}{2017}]{Abbott2017}
{Abbott} B.~P.,  et~al., 2017, \mn@doi [\apjl] {10.3847/2041-8213/aa91c9},
  \href {https://ui.adsabs.harvard.edu/abs/2017ApJ...848L..12A} {848, L12}

\bibitem[\protect\citeauthoryear{{Arcavi} et~al.,}{{Arcavi}
  et~al.}{2017}]{Arcavi2017}
{Arcavi} I.,  et~al., 2017, \mn@doi [\nat] {10.1038/nature24291}, \href
  {https://ui.adsabs.harvard.edu/abs/2017Natur.551...64A} {551, 64}

\bibitem[\protect\citeauthoryear{{Chornock} et~al.,}{{Chornock}
  et~al.}{2017}]{Chornock2017}
{Chornock} R.,  et~al., 2017, \mn@doi [\apjl] {10.3847/2041-8213/aa905c}, \href
  {https://ui.adsabs.harvard.edu/abs/2017ApJ...848L..19C} {848, L19}

\bibitem[\protect\citeauthoryear{{Coulter} et~al.,}{{Coulter}
  et~al.}{2017}]{Coulter2017}
{Coulter} D.~A.,  et~al., 2017, \mn@doi [Science] {10.1126/science.aap9811},
  \href {https://ui.adsabs.harvard.edu/abs/2017Sci...358.1556C} {358, 1556}

\bibitem[\protect\citeauthoryear{{Cowan}, {Sneden}, {Lawler}, {Aprahamian},
  {Wiescher}, {Langanke}, {Mart{\'\i}nez-Pinedo}  \& {Thielemann}}{{Cowan}
  et~al.}{2021}]{Cowan2021}
{Cowan} J.~J.,  {Sneden} C.,  {Lawler} J.~E.,  {Aprahamian} A.,  {Wiescher} M.,
   {Langanke} K.,  {Mart{\'\i}nez-Pinedo} G.,   {Thielemann} F.-K.,  2021,
  \mn@doi [Reviews of Modern Physics] {10.1103/RevModPhys.93.015002}, \href
  {https://ui.adsabs.harvard.edu/abs/2021RvMP...93a5002C} {93, 015002}

\bibitem[\protect\citeauthoryear{{Cowperthwaite} et~al.,}{{Cowperthwaite}
  et~al.}{2017}]{Cowperthwaite2017}
{Cowperthwaite} P.~S.,  et~al., 2017, \mn@doi [\apjl]
  {10.3847/2041-8213/aa8fc7}, \href
  {https://ui.adsabs.harvard.edu/abs/2017ApJ...848L..17C} {848, L17}

\bibitem[\protect\citeauthoryear{{Cyburt} et~al.,}{{Cyburt}
  et~al.}{2010}]{Cyburt2010}
{Cyburt} R.~H.,  et~al., 2010, \mn@doi [\apjs] {10.1088/0067-0049/189/1/240},
  \href {https://ui.adsabs.harvard.edu/abs/2010ApJS..189..240C} {189, 240}

\bibitem[\protect\citeauthoryear{{Drout} et~al.,}{{Drout}
  et~al.}{2017}]{Drout2017}
{Drout} M.~R.,  et~al., 2017, \mn@doi [Science] {10.1126/science.aaq0049},
  \href {https://ui.adsabs.harvard.edu/abs/2017Sci...358.1570D} {358, 1570}

\bibitem[\protect\citeauthoryear{{Evans} et~al.,}{{Evans}
  et~al.}{2017}]{Evans2017}
{Evans} P.~A.,  et~al., 2017, \mn@doi [Science] {10.1126/science.aap9580},
  \href {https://ui.adsabs.harvard.edu/abs/2017Sci...358.1565E} {358, 1565}

\bibitem[\protect\citeauthoryear{{Fahlman} \& {Fern{\'a}ndez}}{{Fahlman} \&
  {Fern{\'a}ndez}}{2018}]{Fahlman2018}
{Fahlman} S.,  {Fern{\'a}ndez} R.,  2018, \mn@doi [\apjl]
  {10.3847/2041-8213/aaf1ab}, \href
  {https://ui.adsabs.harvard.edu/abs/2018ApJ...869L...3F} {869, L3}

\bibitem[\protect\citeauthoryear{{Fuller}, {Fowler}  \& {Newman}}{{Fuller}
  et~al.}{1982}]{Fuller1982}
{Fuller} G.~M.,  {Fowler} W.~A.,   {Newman} M.~J.,  1982, \mn@doi [\apjs]
  {10.1086/190779}, \href
  {https://ui.adsabs.harvard.edu/abs/1982ApJS...48..279F} {48, 279}

\bibitem[\protect\citeauthoryear{{Goriely}, {Bauswein}  \& {Janka}}{{Goriely}
  et~al.}{2011}]{Goriely2011}
{Goriely} S.,  {Bauswein} A.,   {Janka} H.-T.,  2011, \mn@doi [\apjl]
  {10.1088/2041-8205/738/2/L32}, \href
  {https://ui.adsabs.harvard.edu/abs/2011ApJ...738L..32G} {738, L32}

\bibitem[\protect\citeauthoryear{{Kasen}, {Badnell}  \& {Barnes}}{{Kasen}
  et~al.}{2013}]{Kasen2013}
{Kasen} D.,  {Badnell} N.~R.,   {Barnes} J.,  2013, \mn@doi [\apj]
  {10.1088/0004-637X/774/1/25}, \href
  {https://ui.adsabs.harvard.edu/abs/2013ApJ...774...25K} {774, 25}

\bibitem[\protect\citeauthoryear{{Kasen}, {Fern{\'a}ndez}  \&
  {Metzger}}{{Kasen} et~al.}{2015}]{Kasen2015}
{Kasen} D.,  {Fern{\'a}ndez} R.,   {Metzger} B.~D.,  2015, \mn@doi [\mnras]
  {10.1093/mnras/stv721}, \href
  {https://ui.adsabs.harvard.edu/abs/2015MNRAS.450.1777K} {450, 1777}

\bibitem[\protect\citeauthoryear{{Kasen}, {Metzger}, {Barnes}, {Quataert}  \&
  {Ramirez-Ruiz}}{{Kasen} et~al.}{2017}]{Kasen2017}
{Kasen} D.,  {Metzger} B.,  {Barnes} J.,  {Quataert} E.,   {Ramirez-Ruiz} E.,
  2017, \mn@doi [\nat] {10.1038/nature24453}, \href
  {https://ui.adsabs.harvard.edu/abs/2017Natur.551...80K} {551, 80}

\bibitem[\protect\citeauthoryear{{Korobkin}, {Rosswog}, {Arcones}  \&
  {Winteler}}{{Korobkin} et~al.}{2012}]{Korobkin2012}
{Korobkin} O.,  {Rosswog} S.,  {Arcones} A.,   {Winteler} C.,  2012, \mn@doi
  [\mnras] {10.1111/j.1365-2966.2012.21859.x}, \href
  {https://ui.adsabs.harvard.edu/abs/2012MNRAS.426.1940K} {426, 1940}

\bibitem[\protect\citeauthoryear{{Langanke} \&
  {Mart{\'\i}nez-Pinedo}}{{Langanke} \&
  {Mart{\'\i}nez-Pinedo}}{2000}]{Langanke2000}
{Langanke} K.,  {Mart{\'\i}nez-Pinedo} G.,  2000, \mn@doi [\nphysa]
  {10.1016/S0375-9474(00)00131-7}, \href
  {https://ui.adsabs.harvard.edu/abs/2000NuPhA.673..481L} {673, 481}

\bibitem[\protect\citeauthoryear{{Lattimer} \& {Swesty}}{{Lattimer} \&
  {Swesty}}{1991}]{Lattimer1991}
{Lattimer} J.~M.,  {Swesty} D.~F.,  1991, \mn@doi [\nphysa]
  {10.1016/0375-9474(91)90452-C}, \href
  {https://ui.adsabs.harvard.edu/abs/1991NuPhA.535..331L} {535, 331}

\bibitem[\protect\citeauthoryear{{Lattimer}, {Mackie}, {Ravenhall}  \&
  {Schramm}}{{Lattimer} et~al.}{1977}]{Lattimer1977}
{Lattimer} J.~M.,  {Mackie} F.,  {Ravenhall} D.~G.,   {Schramm} D.~N.,  1977,
  \mn@doi [\apj] {10.1086/155148}, \href
  {https://ui.adsabs.harvard.edu/abs/1977ApJ...213..225L} {213, 225}

\bibitem[\protect\citeauthoryear{{Li} \& {Paczy{\'n}ski}}{{Li} \&
  {Paczy{\'n}ski}}{1998}]{LiandPac1998}
{Li} L.-X.,  {Paczy{\'n}ski} B.,  1998, \mn@doi [\apjl] {10.1086/311680}, \href
  {https://ui.adsabs.harvard.edu/abs/1998ApJ...507L..59L} {507, L59}

\bibitem[\protect\citeauthoryear{{Lippuner} \& {Roberts}}{{Lippuner} \&
  {Roberts}}{2017}]{Lippuner2017}
{Lippuner} J.,  {Roberts} L.~F.,  2017, \mn@doi [\apjs]
  {10.3847/1538-4365/aa94cb}, \href
  {https://ui.adsabs.harvard.edu/abs/2017ApJS..233...18L} {233, 18}

\bibitem[\protect\citeauthoryear{{McCully} et~al.,}{{McCully}
  et~al.}{2017}]{McCully2017}
{McCully} C.,  et~al., 2017, \mn@doi [\apjl] {10.3847/2041-8213/aa9111}, \href
  {https://ui.adsabs.harvard.edu/abs/2017ApJ...848L..32M} {848, L32}

\bibitem[\protect\citeauthoryear{{Metzger} et~al.,}{{Metzger}
  et~al.}{2010}]{Metzger2010}
{Metzger} B.~D.,  et~al., 2010, \mn@doi [\mnras]
  {10.1111/j.1365-2966.2010.16864.x}, \href
  {https://ui.adsabs.harvard.edu/abs/2010MNRAS.406.2650M} {406, 2650}

\bibitem[\protect\citeauthoryear{{Metzger}, {Thompson}  \&
  {Quataert}}{{Metzger} et~al.}{2018}]{Metzger2018}
{Metzger} B.~D.,  {Thompson} T.~A.,   {Quataert} E.,  2018, \mn@doi [\apj]
  {10.3847/1538-4357/aab095}, \href
  {https://ui.adsabs.harvard.edu/abs/2018ApJ...856..101M} {856, 101}

\bibitem[\protect\citeauthoryear{{Meyer}}{{Meyer}}{1989}]{Meyer1989}
{Meyer} B.~S.,  1989, \mn@doi [\apj] {10.1086/167702}, \href
  {https://ui.adsabs.harvard.edu/abs/1989ApJ...343..254M} {343, 254}

\bibitem[\protect\citeauthoryear{{Morozova}, {Piro}, {Renzo}, {Ott}, {Clausen},
  {Couch}, {Ellis}  \& {Roberts}}{{Morozova} et~al.}{2015}]{Morozova2015}
{Morozova} V.,  {Piro} A.~L.,  {Renzo} M.,  {Ott} C.~D.,  {Clausen} D.,
  {Couch} S.~M.,  {Ellis} J.,   {Roberts} L.~F.,  2015, \mn@doi [\apj]
  {10.1088/0004-637X/814/1/63}, \href
  {https://ui.adsabs.harvard.edu/abs/2015ApJ...814...63M} {814, 63}

\bibitem[\protect\citeauthoryear{{M{\"o}sta}, {Radice}, {Haas}, {Schnetter}  \&
  {Bernuzzi}}{{M{\"o}sta} et~al.}{2020}]{Moesta2020}
{M{\"o}sta} P.,  {Radice} D.,  {Haas} R.,  {Schnetter} E.,   {Bernuzzi} S.,
  2020, \mn@doi [\apjl] {10.3847/2041-8213/abb6ef}, \href
  {https://ui.adsabs.harvard.edu/abs/2020ApJ...901L..37M} {901, L37}

\bibitem[\protect\citeauthoryear{{Nativi}, {Bulla}, {Rosswog}, {Lundman},
  {Kowal}, {Gizzi}, {Lamb}  \& {Perego}}{{Nativi} et~al.}{2021}]{Nativi2021}
{Nativi} L.,  {Bulla} M.,  {Rosswog} S.,  {Lundman} C.,  {Kowal} G.,  {Gizzi}
  D.,  {Lamb} G.~P.,   {Perego} A.,  2021, \mn@doi [\mnras]
  {10.1093/mnras/staa3337}, \href
  {https://ui.adsabs.harvard.edu/abs/2021MNRAS.500.1772N} {500, 1772}

\bibitem[\protect\citeauthoryear{{Nedora}, {Bernuzzi}, {Radice}, {Perego},
  {Endrizzi}  \& {Ortiz}}{{Nedora} et~al.}{2019}]{Nedora2019}
{Nedora} V.,  {Bernuzzi} S.,  {Radice} D.,  {Perego} A.,  {Endrizzi} A.,
  {Ortiz} N.,  2019, \mn@doi [\apjl] {10.3847/2041-8213/ab5794}, \href
  {https://ui.adsabs.harvard.edu/abs/2019ApJ...886L..30N} {886, L30}

\bibitem[\protect\citeauthoryear{{Nicholl} et~al.,}{{Nicholl}
  et~al.}{2017}]{Nicholl2017}
{Nicholl} M.,  et~al., 2017, \mn@doi [\apjl] {10.3847/2041-8213/aa9029}, \href
  {https://ui.adsabs.harvard.edu/abs/2017ApJ...848L..18N} {848, L18}

\bibitem[\protect\citeauthoryear{{O'Connor} \& {Ott}}{{O'Connor} \&
  {Ott}}{2010}]{OConnor2010}
{O'Connor} E.,  {Ott} C.~D.,  2010, \mn@doi [Classical and Quantum Gravity]
  {10.1088/0264-9381/27/11/114103}, \href
  {https://ui.adsabs.harvard.edu/abs/2010CQGra..27k4103O} {27, 114103}

\bibitem[\protect\citeauthoryear{{Oda}, {Hino}, {Muto}, {Takahara}  \&
  {Sato}}{{Oda} et~al.}{1994}]{Oda1994}
{Oda} T.,  {Hino} M.,  {Muto} K.,  {Takahara} M.,   {Sato} K.,  1994, \mn@doi
  [Atomic Data and Nuclear Data Tables] {10.1006/adnd.1994.1007}, \href
  {https://ui.adsabs.harvard.edu/abs/1994ADNDT..56..231O} {56, 231}

\bibitem[\protect\citeauthoryear{{Ott} et~al.,}{{Ott} et~al.}{2013}]{Ott2013}
{Ott} C.~D.,  et~al., 2013, \mn@doi [\apj] {10.1088/0004-637X/768/2/115}, \href
  {https://ui.adsabs.harvard.edu/abs/2013ApJ...768..115O} {768, 115}

\bibitem[\protect\citeauthoryear{{Perego} et~al.,}{{Perego}
  et~al.}{2020}]{Perego2020}
{Perego} A.,  et~al., 2020, arXiv e-prints, \href
  {https://ui.adsabs.harvard.edu/abs/2020arXiv200908988P} {p. arXiv:2009.08988}

\bibitem[\protect\citeauthoryear{{Pian} et~al.,}{{Pian}
  et~al.}{2017}]{Pian2017}
{Pian} E.,  et~al., 2017, \mn@doi [\nat] {10.1038/nature24298}, \href
  {https://ui.adsabs.harvard.edu/abs/2017Natur.551...67P} {551, 67}

\bibitem[\protect\citeauthoryear{{Qian} \& {Woosley}}{{Qian} \&
  {Woosley}}{1996}]{Qian1996}
{Qian} Y.~Z.,  {Woosley} S.~E.,  1996, \mn@doi [\apj] {10.1086/177973}, \href
  {https://ui.adsabs.harvard.edu/abs/1996ApJ...471..331Q} {471, 331}

\bibitem[\protect\citeauthoryear{{Radice}, {Perego}, {Hotokezaka}, {Fromm},
  {Bernuzzi}  \& {Roberts}}{{Radice} et~al.}{2018}]{Radice2018}
{Radice} D.,  {Perego} A.,  {Hotokezaka} K.,  {Fromm} S.~A.,  {Bernuzzi} S.,
  {Roberts} L.~F.,  2018, \mn@doi [\apj] {10.3847/1538-4357/aaf054}, \href
  {https://ui.adsabs.harvard.edu/abs/2018ApJ...869..130R} {869, 130}

\bibitem[\protect\citeauthoryear{{Soares-Santos} et~al.,}{{Soares-Santos}
  et~al.}{2017}]{SoaresSantos2017}
{Soares-Santos} M.,  et~al., 2017, \mn@doi [\apjl] {10.3847/2041-8213/aa9059},
  \href {https://ui.adsabs.harvard.edu/abs/2017ApJ...848L..16S} {848, L16}

\bibitem[\protect\citeauthoryear{{Symbalisty} \& {Schramm}}{{Symbalisty} \&
  {Schramm}}{1982}]{Symbalisty1982}
{Symbalisty} E.,  {Schramm} D.~N.,  1982, \aplett, \href
  {https://ui.adsabs.harvard.edu/abs/1982ApL....22..143S} {22, 143}

\bibitem[\protect\citeauthoryear{{Tanaka} et~al.,}{{Tanaka}
  et~al.}{2018}]{Tanaka2018}
{Tanaka} M.,  et~al., 2018, \mn@doi [\apj] {10.3847/1538-4357/aaa0cb}, \href
  {https://ui.adsabs.harvard.edu/abs/2018ApJ...852..109T} {852, 109}

\bibitem[\protect\citeauthoryear{{Tanvir} et~al.,}{{Tanvir}
  et~al.}{2017}]{Tanvir2017}
{Tanvir} N.~R.,  et~al., 2017, \mn@doi [\apjl] {10.3847/2041-8213/aa90b6},
  \href {https://ui.adsabs.harvard.edu/abs/2017ApJ...848L..27T} {848, L27}

\bibitem[\protect\citeauthoryear{{Villar} et~al.,}{{Villar}
  et~al.}{2017}]{Villar2017}
{Villar} V.~A.,  et~al., 2017, \mn@doi [\apjl] {10.3847/2041-8213/aa9c84},
  \href {https://ui.adsabs.harvard.edu/abs/2017ApJ...851L..21V} {851, L21}

\bibitem[\protect\citeauthoryear{{Wollaeger} et~al.,}{{Wollaeger}
  et~al.}{2018}]{Wollaeger2018}
{Wollaeger} R.~T.,  et~al., 2018, \mn@doi [\mnras] {10.1093/mnras/sty1018},
  \href {https://ui.adsabs.harvard.edu/abs/2018MNRAS.478.3298W} {478, 3298}

\bibitem[\protect\citeauthoryear{{Wu}, {Ricigliano}, {Kashyap}, {Perego}  \&
  {Radice}}{{Wu} et~al.}{2021}]{Wu2021}
{Wu} Z.,  {Ricigliano} G.,  {Kashyap} R.,  {Perego} A.,   {Radice} D.,  2021,
  arXiv e-prints, \href {https://ui.adsabs.harvard.edu/abs/2021arXiv211106870W}
  {p. arXiv:2111.06870}

\makeatother
\end{thebibliography}








\bsp	
\label{lastpage}
\end{document}